\documentclass[twocolumn]{aastex63}
\usepackage{array}
\usepackage{multirow}
\usepackage{xcolor} %Only needed for the note commands I added.
\usepackage{amsmath}

 \received{August 25, 2022}
 \accepted{September 16, 2022}
\shorttitle{On large-scale dynamos with stable stratification}
\shortauthors{V. Skoutnev, J. Squire, A. Bhattacharjee}
\graphicspath{{./}{}}
\begin{document}

\title{On large-scale dynamos with stable stratification and the application to stellar radiative zones}

\correspondingauthor{Valentin Skoutnev}
\email{valentinskoutnev@gmail.com}

\author{V. Skoutnev}
 \affiliation{Department of Astrophysical Sciences and Max Planck Princeton Center, Princeton University, Princeton, NJ 08544, USA}
\author{J. Squire}
\affiliation{Physics Department, University of Otago, Dunedin 9010, New Zealand}
\author{A. Bhattacharjee}
\affiliation{Department of Astrophysical Sciences, Princeton University, Princeton, NJ 08544, USA}




\begin{abstract}

Our understanding of large-scale magnetic fields in stellar radiative zones remains fragmented and incomplete. Such magnetic fields, which must be produced by some form of dynamo mechanism, are thought to dominate angular-momentum transport, making them crucial to stellar evolution. A major difficulty is the effect of stable stratification, which generally suppresses dynamo action. We explore the effects of stable stratification on mean-field dynamo theory with a particular focus on a non-helical large-scale dynamo (LSD) mechanism known as the magnetic shear-current effect. We find that the mechanism is robust to increasing stable stratification as long as the original requirements for its operation are met: a source of shear and non-helical magnetic fluctuations (e.g. from a small-scale dynamo). Both are plausibly sourced in the presence of differential rotation. Our idealized direct numerical simulations, supported by mean-field theory, demonstrate the generation of near equipartition large-scale toroidal fields.  Additionally, a scan over magnetic Reynolds number shows no change in the growth or saturation of the LSD, providing good numerical evidence of a dynamo mechanism resilient to catastrophic quenching, which has been an issue for helical dynamos. These properties---the absence of catastrophic quenching and robustness to stable stratification---make the mechanism a plausible candidate for generating in-situ large-scale magnetic fields in stellar radiative zones. 
\end{abstract}

%%%%%%%%%%%%%%%%%%%%%%%%%%%%%%%%%%%%%%%%%%%%%%%%%%

%%%%%%%%%%%%%%%%% BODY OF PAPER %%%%%%%%%%%%%%%%%%
\section{Introduction}
Massive stars are born rapidly rotating and quantifying their spin down throughout the course of stellar evolution is a key issue in stellar astrophysics. The angular momentum present in the radiative interior at the end of a massive star's life has a strong impact on the dynamics of core-collapse and the spin distribution at formation of the subsequent compact remnant \citep{macfadyen1999,heger2000,yoon2006}. Measurements of core rotation rates of red giant stars \citep{Cantiello_2014,eggenberger2017,ouazzani2019}, white dwarf spins \citep{hermes2017}, rotation periods of neutron stars at birth \citep{faucher2006,gullon2014}, and black hole spins by LIGO and Virgo Collaborations \citep{zaldarriaga2018,roulet2021} all indicate efficient angular momentum transport in the progenitor cores that cannot be explained generally by hydrodynamic processes. Instead, torques from large-scale magnetic fields generated by a dynamo are often invoked as the dominant form of angular momentum transport in regions of radial differential rotation. The leading candidate is the Tayler-Spruit (TS) dynamo \citep{spruit2002}, whose recently updated prescription in 1D stellar evolution codes \citep{fuller2019} finds generally improved agreement with observations \citep{FullerLu2022}, although discrepancies remain \citep{eggenberger2019asteroseismology,den2020asteroseismology}. First-principles investigations of the TS dynamo are pressingly needed to inform 1D prescriptions, similar to the feedback between mixing-length theory and 3D convection simulations in the modeling of stellar convection zones. In particular, we still lack understanding of what nonlinear dynamo mechanisms could enable the TS mechanism, remaining viable in the high magnetic Reynolds number and stably stratified conditions of a radiative zone (RZ).

In the global context, the Tayler-Spruit  dynamo provides torques through the Maxwell stress of axisymmetric poloidal and toroidal magnetic fields  whose  energy  is  sourced  from  the radial differential  rotation (DR) itself.  Successful  operation requires closure of a dynamo loop: the axisymmetric toroidal field needs to be generated from the axisymmetric poloidal field and vice versa. The first direction is straightforward and uncontroversial: the toroidal field is generated by winding of the axisymmetric poloidal  field  by  the radial DR.  However,  the  regeneration  of  the  axisymmetric poloidal  field,  remains  an  open  question  and  is  likely  a  non-linear dynamo effect \citep{zahn2007,fuller2019}. The original \citet{spruit2002} study suggested that the amplified toroidal field goes unstable to the Tayler instability \citep{Tayler1973,markey1973adiabatic} and creates poloidal field but, as first pointed out by \citet{zahn2007},  this is insufficient because only non-axisymmetric modes are generated by the Tayler instability. 
Alternatively, follow-up studies proposed that a non-linear dynamo effect such as the $\alpha$-effect (driven by the helical part of the turbulence from the Tayler instability) may close the dynamo loop \citep{zahn2007,fuller2019}.  However, the alpha effect seems to suffer catastrophic quenching at high magnetic Reynolds numbers in the presence of small-scale magnetic fields, which means that the large-scale dynamo might saturate on resistive timescales \citep{cattaneo1996nonlinear,brandenburg2001inverse,rincon2019} (see \citet{brandenburg2018advances} and references within on possible ways to avoid quenching). The timescales of resistive diffusion  can be comparable to or longer than stellar lifetimes. An alternative dynamo mechanism that is immune to quenching and can generate equipartition large-scale magnetic fields on dynamical timescales is thus desirable.

Such a non-linear dynamo mechanism must also be able to operate in the stably stratified conditions characteristic of stellar RZs. Stable stratification generally tends to suppress dynamo action because it imposes a particular form of anisotropy on the velocity field: fluid motions are unrestricted horizontally while vertical motions are rapidly restored \citep{Riley_2000,billant2001self,lindborg2006,brethouwer2007,chini_2022}. In the limit of arbitrarily strong stable stratification, the velocity field is approximated by a two-component and three-dimensional field, which is well known to inhibit dynamo action \citep{Zeldovich1980}. Thus stable stratification and DR tend to play opposite roles in suppressing and supporting dynamo action. Finding a successful dynamo loop driven by DR while surviving the extreme stable stratification of RZs is a primary challenge of stellar interior physics.

One promising mechanism is the magnetic shear-current (MSC) effect, in which large-scale magnetic fields result from the combination of mean shear and non-helical magnetic fluctuations \citep{rogachevskii2004nonlinear,squire2015generation,squire2015electromotive,Squire2016}. Originally applied in the context of accretion disks, the MSC effect helps explain large-scale magnetic field generation in magnetic turbulence driven by the magneto-rotational instability in a sheared (Keplerian) flow \citep{lesur2008self,squire2015statistical,shi2016saturation}. In the stellar context, we propose that the MSC effect would operate as follows: a radial shear generates a toroidal field from the poloidal field, whereupon statistical correlations in the non-helical magnetic turbulence lead to an off-diagonal turbulent resistivity that sources poloidal field from the toroidal field, thus closing the dynamo loop. The magnetic fluctuations can in principle originate from a variety of sources including magnetic turbulence from the Tayler instability \citep{zahn2007,fuller2019} or from a small-scale dynamo (SSD) operating in stably stratified turbulence \citep{Skoutnev_2021} driven by hydrodynamic instabilities, such as horizontal shear instabilities of latitudinal DR \citep{zahn1974rotational,zahn1992circulation,prat2013turbulent,prat2014shear,cope2020dynamics,garaud2020horizontal,garaudJtCS2021}. Additionally, because the MSC effect is driven by the non-helical part of the magnetic turbulence, the generated large-scale fields are also non-helical and are not subject to MHD helicity constraints\footnote{Strictly speaking, the MSC effect generates purely non-helical fields only for periodic boundary conditions or for fields that vanish sufficiently fast outside a finite domain. Closed boundary conditions could lead to a shear-current effect that generates helicity \citep{brandenburg2005BC_helicityflux}.}. These constraints can lead to so-called ``catastrophic quenching", whereby LSD saturation occurs at an amplitude, or on a timescale, that prohibitively scales with the microscopic diffusion coefficients \citep{gruzinov1994self,bhattacharjee1995self,rogachevskii2004nonlinear,rincon2019}. An extremely inefficient LSD results in the astrophysical limit of large $Rm$ unless helicity fluxes through the boundaries of the system are sufficiently large \citep{blackman2000constraints,vishniac2001magnetic,kleeorin2000helicity,brandenburg2002magnetichelicity}. Because of their non-helical nature, MSC driven dynamos are unlikely to be affected by catastrophic quenching in the same way as $\alpha$ dynamos driven by helical turbulence. This property, in combination with ample sources of magnetic turbulence and shear flows, make the MSC effect a promising mechanism for locally generating large-scale magnetic fields  without restrictions imposed by the level of helicity fluxes in differentially rotating stellar RZs \citep{kissin2018rotation}.

The aim of this paper is to extend mean-field theory to include the effects of stable stratification and subsequently assess the viability of the MSC effect as a dynamo mechanism. This is an important step for a better understanding of large-scale magnetic field generation in the stellar context. The primary issue is the generally unknown effect of stable stratification on large-scale dynamo mechanisms, particularly in the extreme parameter regimes of high magnetic Reynolds number and strong stable stratification. Can the large-scale dynamo operate in the background of weak stable stratification? What level of stratification is needed to shut down a large-scale dynamo mechanism? Where in parameter space are stellar RZs relative to this threshold? We attempt to answer these questions with combined analytical and numerical approaches. We use an available analytical framework (mean-field theory) to study perturbatively the effects of weak stratification on large-scale dynamos followed by direct numerical simulations (DNS) to additionally study the non-perturbative limit of strong stratification. A general agreement between analytic and numerical results allows us to extrapolate these results to realistic parameters for stellar RZs.

\subsection{Paper Outline} 

Section \ref{sec:Theory} presents our local model of a stellar RZ and a mean-field theory framework for the MHD Boussinesq system. This allows perturbative calculation of the modifications to the MSC effect that result from stratification. However, the non-perturbative limit of strong stable stratification relevant to stars lies outside the formally valid regime of mean-field theory and therefore requires numerical exploration. Section \ref{sec:Numerical} then presents sets of DNS where the effect of varying the stable stratification on the LSD is examined. Section \ref{sec:Application} discusses two possible applications of the MSC effect in RZs and Section \ref{sec:Conclusion} concludes. 

\section{Theoretical Considerations}\label{sec:Theory}
\subsection{Model of Turbulence in a Radiative Zone}
\begin{figure}
    \centering
    \includegraphics[width=\linewidth]{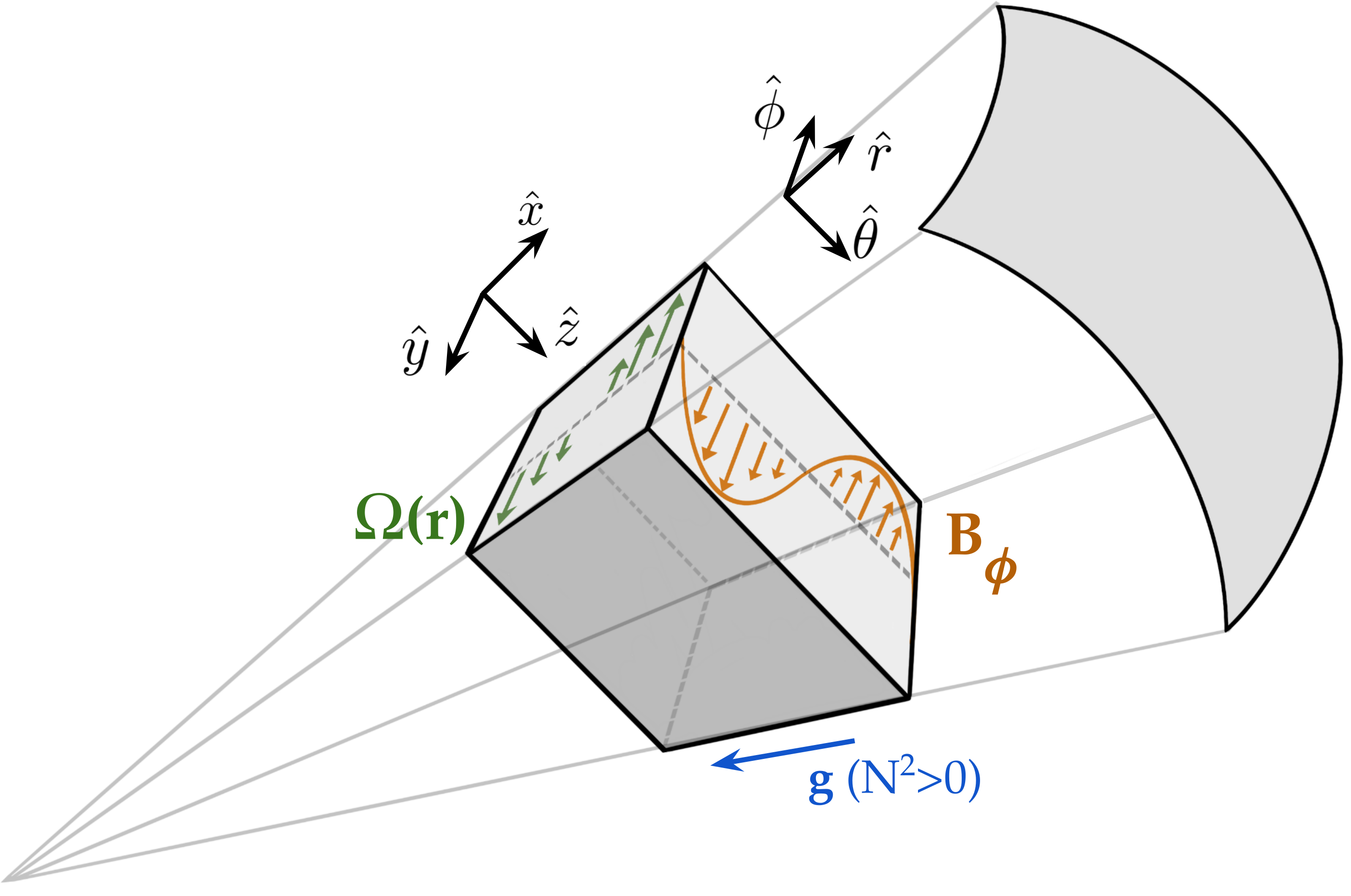}
    \caption{Schematic of a local shearing box embedded in a section of a differentially rotating radiative zone. Orientation of the Cartesian coordinates used in our setup are shown relative to the local unit vectors of the spherical coordinate system ( $\hat{x}=\hat{r}, \hat{y}=-\hat{\phi}, \hat{z}=\hat{\theta}$).}
    \label{fig:Diagram}
\end{figure}
To most simply capture the physics of a local section of a differentially rotating RZ, we consider a shearing box model as shown in Figure \ref{fig:Diagram} with an imposed vertical shear profile, stable stratification, and a turbulent velocity field driven with a body force at intermediate scales. The vertical (radial) shear profile represents a local section of radial DR, $\Omega(r)$, which has a shear rate $S=r(\partial\Omega/\partial r)$ in the rotating frame. The driven turbulence represents stably stratified turbulence that may be sourced from an instability. We ignore mean rotation to focus on understanding the novel effect of stable stratification on the LSD, a critical step to the full problem. We discuss the possible effects of mean rotation in Section \ref{sec:Application}.  

This setup allows a study of the SSD and the LSD as well as their crucial nonlinear feedback. We require that 1) the scale separation between the outer box scale and the intermediate forcing scale is large enough to allow an unambiguous definition of large scales and 2) the scale separation between the intermediate forcing scale and the smallest (viscous) scales is large enough to allow a significant turbulent cascade.

The unstratified, non-rotating case of this setup is actually destabilized by a large scale hydrodynamic instability known as the vorticity dynamo (VD) \citep{Elperin2003,PKapyla2009}. The VD generates large-scale vortical (shear) flows and saturates at large amplitudes that may disrupt the operation of a LSD \citep{teed2016}, although it can be suppressed by sufficient rotation \citep{kapyla2022compressible}. Therefore, it is important to understand the complications caused by the VD. We work through the mean-field framework for both the LSD and VD in the next sections. It will turn out that the VD is also strongly suppressed by stable stratification and so does not play a role in the stratified problem. 

\subsection{Equations}
The standard hydrodynamical model of a stably stratified, collisional plasma with subsonic velocity fluctuations on vertical length scales that are small compared to the local scale-height is the set of Spiegel–Veronis–Boussinesq equations \citep{spiegelBouss}. With magnetic fields included, we will call them the MHD Boussinesq equations. In the following sections, we apply the mean-field approach to the MHD Boussinesq equations for the total fields to separate the evolution of the large and small-scale fields. This allows for an analysis of the two large-scale instabilities of the system: the hydrodynamic vorticity dynamo of the velocity field and the large-scale dynamo of the magnetic field. 

The equations for the total quantities (the velocity field $\textbf{U}_T$, magnetic field $\textbf{B}_T$, buoyancy $\Theta_T$, and pressure $P_T$) are:
\begin{align}
\partial_t {\textbf{U}_T}+{\textbf{U}_T}\cdot \nabla {\textbf{U}_T}=-\nabla {P_T}+\Theta_T\hat{x}+\textbf{J}_T\times\textbf{B}_T+\nu\nabla^2 \textbf{U}_T+\sigma_f,\nonumber
\end{align}

\begin{align}
 \partial_t \textbf{B}_T =\nabla \times(\textbf{U}_T\times \textbf{B}_T)+\eta\nabla^2 \textbf{B}_T,
\end{align}

\begin{equation}
\partial_t \Theta_T+\textbf{U}_T\cdot\nabla \Theta_T=-N^2\textbf{U}_{T}\cdot\hat{x}+\kappa\nabla^2\Theta_T,
\end{equation}

\begin{equation}
\nabla \cdot \textbf{U}_T=0, \: \nabla \cdot \textbf{B}_T=0,
\end{equation}
where $\nu$ is the kinematic viscosisty, $\eta$ is the resistivity, $\kappa$ is the thermal diffusivity, $\sigma_f$ is non-helical forcing at small scales, and $\textbf{B}_T$ is normalized by $\sqrt{4\pi\rho_0}$ ($\rho_0$ is the constant background plasma density). Note that the "vertical"  direction is aligned with $\hat{x}$. In this formulation \citep{spiegelBouss,kundu2002fluid,garaudJtCS2021}, the buoyancy field is related to the temperature fluctuations by $\Theta_T=\alpha_V g T'_T$ and the Brunt-V$\ddot{\mathrm{a}}$is$\ddot{\mathrm{a}}$l$\ddot{\mathrm{a}}$ frequency is given by $N^2=\alpha_V g(T_{0,x}-T_{\mathrm{ad},x})>0$, where $g>0$ is the local gravitational acceleration, $\alpha_V$ is the coefficient of thermal expansion, and $T_{0,x}$ and $T_{\mathrm{ad},x}$ are the background and adiabatic temperature gradients, respectively.

The mean-field approach splits the total fields into mean and fluctuating components, which we will denote with upper and lower case letters, respectively. We define the mean field as a spatial average over $x$ and $y$ (azimuthal and radial average) of the total field (e.g. $\langle \textbf{B}_T \rangle=\int\int \textbf{B}_Tdxdy=\textbf{B}(z)$). Thus the magnetic field is $\textbf{B}_T=\textbf{B}+\textbf{b}$, the velocity field is $\textbf{U}_T=\textbf{U}'+\textbf{u}$ with $\textbf{U}'=\textbf{U}_0+\textbf{U}$, and the buoyancy field is $\Theta_T=\Theta+\theta$. Note that the divergence-free conditions require $B_z(z)$ and $U_z(z)$ to be constants (which we set to $0$), so the mean magnetic field, for example, is of the form $\textbf{B}(z,t)=(B_x(z,t),B_y(z,t),0)$. The local radial DR is modeled as an imposed linear shear flow $\textbf{U}_0=-Sx\hat{y}$, which varies vertically and flows in the "toroidal" $\hat{y}$ direction. See Figure \ref{fig:Diagram} for the geometry. Lastly, to aid with studying the VD, we define the vorticity $\textbf{W}_T=\nabla \times \textbf{U}_T=\textbf{W}'+\textbf{w}$, where $\textbf{W}'=\textbf{W}_0+\textbf{W}$ and $\textbf{W}_0=\nabla \times \textbf{U}_0=-S\hat{z}$.

The mean-field equations are obtained by substituting the scale-separated fields into the equations for the total quantities and taking spatial averages:

\begin{align}
\partial_t {\textbf{W}}=&\nabla\times\left(\textbf{U}' \times \textbf{W}'+\textbf{J}\times \textbf{B}\right)+\partial_z\Theta\hat{y}\\
&+\nabla\times\left(\nabla \cdot \left(\langle-\textbf{u}  \textbf{u}+\textbf{b}\textbf{b}\rangle\right) \right)+\nu\nabla^2 {\textbf{W}},\nonumber
\end{align}

\begin{equation}
\partial_t {\textbf{B}} =\nabla \times(\textbf{U}'\times {\textbf{B}}+\langle\textbf{u}\times \textbf{b}\rangle)+\eta\nabla^2 {\textbf{B}},
\end{equation}

\begin{equation}
\partial_t \Theta+\textbf{U}'\cdot\nabla \Theta+\langle \textbf{u}\cdot\nabla \theta\rangle=-N^2\textbf{U}\cdot\hat{x}+\kappa\nabla^2\Theta,
\end{equation}

\begin{equation}
\nabla \cdot \textbf{U}=0, \: \nabla \cdot \textbf{B}=0,
\end{equation}

In this form, it is clear that the mean vorticity and  mean magnetic field can be driven by the Reynolds and Maxwell stresses, $\mathbf{\mathcal{T}}=\langle-\textbf{uu}+\textbf{bb}\rangle$, and the electromotive force, $\mathbf{\mathcal{E}}=\langle \textbf{u}\times \textbf{b}\rangle$, respectively, of the fluctuating quantities. The equations for the fluctuating quantities are obtained by subtracting the equations for the mean-fields from the those of the total fields. The fluctuation equations are discussed in Appendix \ref{sec:SOCA} where they are used to compute transport coefficients.

The small scales contain stably stratified turbulence (driven by $\sigma_f$) and the small-scale dynamo. We now discuss their details as well as the dimensionless parameters of the problem before continuing with mean-field theory of the large scales.

\subsection{Dimensionless parameters}
Suppose the forcing $\sigma_f$ at length scale $l_f=2\pi/k_f$ leads to steady-state turbulence with outer-scale velocity fluctuations  $u_{\mathrm{rms}}$ prior to the growth of any instabilities. The system is then described by five dimensionless parameters, $Re$, $Sh$, $Fr$, $Pm$, and $Pr$. The Reynolds number $Re=u_{\mathrm{rms}}/k_f\nu$ is the ratio of the viscous timescale to the outer scale eddy turnover time. The shear number $Sh=Sl_f/ u_{\mathrm{rms}}$ captures the ratio of the outer scale eddy turnover time to the shearing time scale. The Froude number $Fr=u_{\mathrm{rms}}/Nl_f$ is the ratio of the gravitational restoring timescale (Brunt-V$\ddot{\mathrm{a}}$is$\ddot{\mathrm{a}}$l$\ddot{\mathrm{a}}$ period) to the outer scale eddy turnover time. Lastly, the magnetic Prandtl $Pm=\nu/\eta$ and thermal Prandtl $Pr=\nu/\kappa$ numbers measure ratios of diffusivity timescales. We set both to unity $Pm=Pr=1$ for simplicity in our DNS, but will discuss their expected effects based on theory and previous simulations in Section \ref{sec:Application}.

\subsection{Stably Stratified Hydrodynamic Turbulence}
The forcing, $\sigma_f$, leads to stably stratified turbulence that provides the background, hydrodynamic turbulence in which the SSD, VD, and LSD may grow. We briefly review properties relevant to the dynamo. When the $Pr=O(1)$, the turbulent cascade in a stably stratified fluid with energy injection $\epsilon\approx u_{\mathrm{rms}}^3k_f$ at wavenumber $k_f$ and dissipation at viscous wavenumber $k_\nu$ contains two inertial ranges, one above and one below the Ozmidov wavenumber $k_O/2\pi=(N^3/\epsilon)^{1/2}$ where the eddy turnover frequency matches the Brunt-V$\ddot{\mathrm{a}}$is$\ddot{\mathrm{a}}$l$\ddot{\mathrm{a}}$ frequency \citep{ozmidov1992variability,brethouwer2007}. At larger scales with wavenumbers less than $k_O$, the velocity field is highly anisotropic due to the restriction of vertical motions by the stable stratification \citep{Riley_2000,riley2010recent}. The scale separation between $k_f$ and $k_O$ is controlled by the Froude number $k_O=Fr^{-3/2}k_f$. At smaller scales with wavenumbers greater than $k_O$, the velocity field is nearly isotropic due to the negligible effect of buoyancy on the fast timescales of small eddies. The scale separation between $k_O$ and $k_\nu$ is set by the buoyancy Reynolds number $k_\nu=Rb^{3/4}k_O$, where $Rb=ReFr^2$. $Rb$ has been found to be the primary control parameter of stably stratified turbulence and needs to be larger than one to avoid the viscosity-affected stratified flow regime where the isotropic inertial range disappears \citep{billant2001self,waite2004stratified,lindborg2006,brethouwer2007,maffioli2016dynamics,chini_2022}. DNS of strong stably stratified turbulence thus simultaneously requires $Fr\ll1$ and $Rb\gg1$, which is computationally challenging \citep{bartello2013sensitivity}.

\subsection{Small-Scale Dynamo}
The SSD typically has a much faster growth rate than either the VD or LSD and generates magnetic fields primarily on length scales smaller than the forcing scale. Here we briefly review the instability criterion. The SSD will operate if the turbulence is sufficiently vigorous: magnetic stretching will statistically win over magnetic diffusion and lead to amplification of any seed magnetic field to near-equipartition with the turbulent kinetic energy. In isotropic turbulence, the SSD is unstable when $Rm$ is above a critical value $Rm>Rm^c=O(10^2)$. $Rm^c$ depends on the magnetic Prandtl number $Rm^c=Rm^c(Pm)$ and is higher in the low $Pm$ limit applicable to stellar interiors \citep{Iskakov_lowpm}. When unstable, its exponential growth rate scales as $\gamma_{\rm SSD}\sim u_{\mathrm{rms}}Re^{1/2}/l_f$ \citep{rincon2019}. However, in the presence of stable stratification ($Fr<1$) with $Pr=O(1)$, the largest scales in the system are anisotropic and inefficient at contributing to the dynamo, leading to a reduced effective $Rm$ known as the magnetic buoyancy Reynolds number $Rb_m=RmFr^2$ \citep{Skoutnev_2021}. The new criterion for the SSD to operate becomes $Rb_m>Rb_m^c$, where $Rb_m^c$ has a dependence on $Pm$ \citep{Skoutnev_2021} similar to $Rm^c$ in isotropic turbulence.  Thus, strong enough stable stratification ($Fr\ll1$) can shut off the SSD even at high magnetic Reynolds numbers $Rm\gg Rm^c$. Note that in the low $Pr\ll1$ limit relevant to stellar interiors, the dynamo is more efficient since increased thermal diffusion reduces the effects of buoyancy (see \cite{skoutnev2022critical} for the modified instability criterion). In this framework, the SSD operates within the fluctuation equations (see Appendix \ref{sec:SOCA}) and rapidly provides a source of background magnetic fluctuations that interact with the LSD.

\subsection{Mean-Field Theory}
At this point, the mean-field and fluctuation equations together are still exact and just as difficult to solve as the original MHD Boussinesq equations. The primary issue is finding a closure for the evolution of the mean vorticity and magnetic fields driven by $\mathbf{\mathcal{T}}$ and $\mathbf{\mathcal{E}}$, respectively. To work around this issue, we use the second order correlation approximation (SOCA), which works with linear fluctuation equations by neglecting the problematic third and higher-order terms \citep{brandenburg2005astrophysical,radler2006mean,squire2015electromotive}. This approximation enables a closed system of equations for the mean-field evolution. A background of isotropic small-scale turbulence in both the velocity and magnetic field is assumed onto which anisotropic effects such as shear and stratification are added perturbatively. Physically, the small-scale magnetic field should arise from the SSD but its statistics are treated as given for this calculation. Arbitrarily small seeds of the mean-fields may then be linearly unstable. 

The choice of horizontal averages leaves the mean-field vorticity and induction equations  (e.g. $\textbf{J}\times \textbf{B}=0$, $\nabla\times(\textbf{U}\times\textbf{B}) =0$) uncoupled except through $\mathbf{\mathcal{E}}$ and $\mathbf{\mathcal{T}}$. We make the standard assumption that $\mathbf{\mathcal{E}}=\mathbf{\mathcal{E}}(\textbf{B})$ depends only on the mean magnetic field, which decouples the mean-field induction and vorticity equations and allows the VD and LSD to be analyzed independently. While it is possible to have joint mean vorticity-magnetic field instabilities \citep{blackman1997vorticity,courvoisier2010self}, simulations in Section \ref{sec:Numerical} support our no-coupling assumption because we always observe the LSD with no accompanying growth of the VD when the VD is suppressed by stratification.

We note that the drastic nature of the SOCA limits the rigorous validity of any results for the LSD to either low magnetic Reynolds numbers $Rm\ll1$ (in the limit of low conductivity $l_f^2/\eta \tau_c\ll 1$) or small Strouhal numbers $St=u_{\mathrm{rms}}\tau_c/l_f\ll1$ (in the limit of high conductivity $l_f^2/\eta \tau_c\gg 1$), where $\tau_c$ is the turbulence correlation time \citep{brandenburg2005astrophysical}. In realistic astrophysical turbulence, $Rm$ is extremely large and $St$ is typically order unity. Results from the SOCA can, as a consequence, be used at most to suggest what effects may be qualitatively operating at $Rm\gg1$ and $St\sim1$. The combination of DNS at moderate $Rm$ alongside the results from mean-field theory is therefore important to improve our confidence in understanding the dynamo mechanisms that operate in astrophysical regimes. 

\subsection{Vorticity Dynamo}
The imposed shear flow in the unstratified case is unstable to a purely hydrodynamic instability known as the vorticity dynamo. We will show that even a small amount of stable stratification in the direction of the imposed mean shear ($\hat{x}$ in our coordinates) will stabilize the VD. We extend the original formulation of the VD in \citet{Elperin2003} to include stable stratification in the framework of mean-field theory. Evolution equations of the mean vorticity field $\textbf{W}(z,t)=(W_x(z,t),W_y(z,t),0)=(-\partial_zU_y,\partial_zU_x,0)$ can be written in the form (dropping magnetic field terms): 

\begin{align}
\partial_t {W_x}&= -SW_y-\nu_{xy}Sl_f^2\partial_z^2W_x+\nu_t u_{\mathrm{rms}}l_f\partial_z^2W_x,\\
\partial_t {W_y}&= -\nu_{yx}Sl_f^2\partial_z^2W_x+\partial_z\Theta+\nu_tu_{\mathrm{rms}}l_f\partial_z^2W_y,\\
\partial_t \Theta&= -N^2U_x+\nu_t u_{\mathrm{rms}}l_f\partial_z^2\Theta.
\end{align}
where $\nu_{xy}$ and $\nu_{yx}$ are the dimensionless off diagonal turbulent viscosities and $\nu_{t}$ is the dimensionless diagonal turbulent viscosity. These equations are identical to that of  \citet{Elperin2003} except for the addition of stratification. We normalize transport coefficients by their expected scalings so that they are dimensionless. Modes of the form $e^{ik_zz+\gamma t}$ have a growth rate:

\begin{equation}\label{eq:VDgrowth}
\gamma_{\mathrm{VD}}=\frac{u_{\mathrm{rms}}}{l_f}\left(\sqrt{-k_z^2l_f^2\nu_{yx}Sh^2-Fr^{-2}}-\nu_t k_z^2l_f^2\right),
\end{equation}
where we use the standard assumptions that the dimensionless transport coefficients are small and there is enough scale separation ($k^2l_f^2|\nu_{xy}|\ll 1$). We see that there are growing solutions for small enough $k_z$, small enough $Fr^{-1}$, and $\nu_{yx}<0$ ($\nu_{yx}$ is negative in the unstratified case \citep{Elperin2003,PKapyla2009}). With the addition of stable stratification, it is clear that even weak stratification reduces or possibly fully stabilizes the growth of the VD since the stratification-related term $Fr^{-2}$ in Eq. \eqref{eq:VDgrowth} is not multiplied by any transport coefficients and $|k_z^2l_f^2\nu_{yx}|\ll 1$. Any modifications of the transport coefficients (e.g. $\nu_{yx}$ or $\nu_t$) by stratification are therefore unimportant because they only appear as higher order corrections to the growth rate.

For a system of arbitrary length in the z-direction, a growing VD with maximum growth rate

\begin{align}
\gamma_{\mathrm{VD}}^{\mathrm{max}}=\frac{u_{\mathrm{rms}}}{l_f}\left(-Sh^2\nu_{yx}/4\nu_t-\nu_tFr^{-2}/Sh^2\nu_{yx}\right),    
\end{align}
occurs at a wavenumber
\begin{align}
k^2_{\mathrm{max}}l_f^2=-Sh^2\nu_{yx}/4\nu_t^2+Fr^{-2}/Sh^2\nu_{yx}.    
\end{align}

Instability requires that the stratification be weaker than $Fr^{-1}< -Sh^2\nu_{yx}/2\nu_t$ or the shear stronger than $Sh>\sqrt{-2\nu_t/Fr\nu_{yx}}\equiv Sh^c_{\mathrm{VD}}$. With the assumption $Sh=O(1)$ and again that the dimensionless transport coefficients are small (and typically off diagonal turbulent diffusivity coefficients are smaller than the diagonals ones $|\nu_{yx}|\ll\nu_t$), even a weak stratification $Fr\approx O(1)$ can shut off the VD. Note that in a finite system, the unstable modes must be able to fit into the domain and so the critical $Sh^c_{\mathrm{VD}}$ for instability may instead depend on the lowest available wavenumber.

\subsection{Large-Scale Dynamo}

In the presence of the imposed shear ($\textbf{U}_0=-Sx\hat{y}$) and non-helical magnetic fluctuations, the system is unstable to a LSD due to the magnetic shear-current effect \citep{Squire2016}. Extending mean-field theory to include stable stratification leads to the evolution equations:

\begin{align}
\partial_t {B_x}&= -\eta_{yx}Sl_f^2\partial_z^2B_y+\eta_t u_{\mathrm{rms}}l_f\partial_z^2B_x,\label{eq:MSCa}\\
\partial_t {B_y}&= -SB_x-\eta_{xy}Sl_f^2\partial_z^2B_x+\eta_t u_{\mathrm{rms}}l_f\partial_z^2B_y.\label{eq:MSCb}
\end{align}
Modes of the form $e^{ik_zz+\gamma t}$  have a growth rate:
\begin{equation}\label{eq:MSCdispersion}
\gamma_{\mathrm{MSC}}=\frac{u_{\mathrm{rms}}}{l_f}\left(k_zl_fSh\sqrt{-\eta_{yx}}-\eta_t k_z^2l_f^2\right),
\end{equation}
where we have used the standard assumption that transport coefficients are small and there is enough scale separation ($k_z^2l_f^2|\eta_{xy}|\ll 1$). There are growing LSD dynamo solutions when the off-diagonal turbulent resistivity is negative $\eta_{yx}<0$, which occurs only when magnetic fluctuations are present \citep{squire2015electromotive}. A positive $\gamma_{\mathrm{MSC}}>0$ requires $Sh>k_{\mathrm{sys}}l_f\eta_t/\sqrt{-\eta_{yx}}\equiv Sh^c_{\mathrm{MSC}}$ for the lowest wavenumber $k_{\mathrm{sys}}$ that fits into the domain. The maximum growth rate and associated wavenumber are given by:

\begin{equation}\label{eq:MSCgammamax}
\gamma_{\mathrm{MSC}}^{\mathrm{max}}=-\frac{u_{\mathrm{rms}}}{l_f}\frac{Sh^2 \eta_{yx}}{4\eta_t},\quad k_{\mathrm{max}}l_f=\frac{Sh\sqrt{-\eta_{yx}}}{2\eta_t}.
\end{equation}

These predictions are consistent with our simulations in Section \ref{sec:IncoherentTest} where we observe that a dominant mode emerges when the domain length $L_z$ is above a critical value and that there are little to no temporal variations in the phase of the growing LSD mode (since the growth rate is purely real).

Unlike for the VD, the mean-field induction equations \eqref{eq:MSCa} and \eqref{eq:MSCb}  look identical to their unstratified case in \citet{Squire2016} except now the the transport coefficients can have additional contributions from the effects of stable stratification. We carry out a calculation of the transport coefficients that incorporates Boussinesq effects into the SOCA framework and report here results relevant to the MSC. Details of the calculation are described in the Appendix \ref{sec:SOCA}. The calculation is general and we also report the effect of stratification on all other transport coefficients in Appendix \ref{sec:SOCA}, including those driven by helical turbulence.

We find that out of $\eta_{yx}$, $\eta_{xy}$, and $\eta_{t}$, only the isotropic turbulent resistivity $\eta_{t}$ is modified by stable stratification as indeed must be the case due to the perturbative expansion in shear and stratification. So now $\eta_{t}=\eta_{t,0}+\eta_{t,N^2}$. The modified coefficient is further split up into contributions from the non-helical velocity and magnetic fluctuations, i.e. $\eta_{t,N^2}=\eta^{(u)}_{t,N^2}+\eta^{(b)}_{t,N^2}$. The result is:

\begin{align}
    \eta^{(u)}_{t,N^2}&=N^2\int dk d\omega \frac{3\tilde{\eta}(\tilde{\nu}\tilde{\kappa}-\omega^2)W_u(k,\omega)}{10(\tilde{\eta}^2+\omega^2)(\tilde{\nu}^2+\omega^2)(\tilde{\kappa}^2+\omega^2)},\\
    \eta^{(b)}_{t,N^2}&=N^2\int dk d\omega\frac{(\tilde{\nu}\tilde{\kappa}^2-(\tilde{\kappa}+2\tilde{\nu})\omega^2)W_b(k,\omega)}{60(\tilde{\nu}^2+\omega^2)^2(\tilde{\kappa}^2+\omega^2)},
\end{align}
where $\tilde{\nu}=\nu k^2$, $\tilde{\eta}=\eta k^2$, $\tilde{\kappa}=\kappa k^2$. $W_u(k,\omega)$ and $W_b(k,\omega)$ are the statistics of the non-helical background velocity and magnetic fluctuations (with the magnetic component assumed to arise from the SSD). We find that for a standard Gaussian model of the fluctuation statistics \citep{radler2006mean} both contributions of stable stratification to $\eta_{t,N^2}$ are positive and that the kinetic term is dominant over the magnetic term (see Appendix \ref{sec:SOCA}). According to Eq. \eqref{eq:MSCgammamax}, mean-field theory predicts that weak stable stratification will slightly weaken the growth rate and push the dominant modes to larger scales (lower $k_{\mathrm{max}}$). In other words, unlike the VD, we expect the LSD to be slightly modified but remain unstable in the presence of stable stratification, so long as there are still sufficient small-scale magnetic fluctuations.

We note that the generated large-scale fields are non-helical under reasonable boundary conditions ( i.e. the total helicity $\mathcal{H}=\int \textbf{B}\cdot \textbf{A} dV=0$, where $\textbf{A}$ is the vector potential). Further discussion of catastrophic quenching and helicity, in particular why standard arguments for catastrophic quenching likely do not apply to the MSC mechanism, is given in Appendix \ref{sec:Helicity}.
\subsection{Summary of Theoretical Predictions}
\subsubsection{Summary without stratification}
If we start with a seed magnetic field, and turn on forcing at $t=0$, then 1) the VD will begin to grow if $Sh>Sh^c_{\mathrm{VD}}$, and 2) the SSD will begin to grow if $Rm>Rm^c$. Once the SSD saturates such that magnetic fluctuation are in equipartition with velocity fluctuations, then the LSD will begin to grow if $Sh>Sh^c_{\mathrm{MSC}}$. Once the mean vorticity and magnetic fields become strong, they can possibly interact with each other through their back reaction on the turbulent flow and its statistics. It turns out in DNS (both in this article at moderate $Re\approx120$ and in \citet{teed2016} at low $Re\approx5$), the zonal shear flow of the VD saturates at amplitudes orders of magnitude larger than the original vertical shear and the driving small-scale turbulence. We interpret this as a destabilization of the model because such a strong zonal flow would likely redistribute its energy in the global context on dynamical timescales and destroy the steady vertical shear assumed in the local box model. Thus we argue it is unphysical to consider the LSD in the context of a local box model in the parameter regimes where the VD is unstable.

\subsubsection{Summary with stable stratification}
With stable stratification that is sufficiently weak, so $Rb_m=RmFr^2>Rb_m^c$ and $Fr\sim 1$, at $t=0$ we expect at least the SSD to grow. The mean-field model predicts a regime where the VD will be stable ($Sh<Sh^c_{\mathrm{VD}}$), but the LSD will continue to operate if $Sh>Sh^c_{\mathrm{MSC}}$. This is astrophysically interesting since it allows the in-situ generation of a mean magnetic field without destabilization of the background hydrodynamic flow. Since stable stratification is only added perturbatively in the SOCA, it cannot predict how the LSD will behave with increasingly stronger stratification (increasing $Fr^{-1}$). Further increasing the stable stratification will slowly suppress the SSD and also increase $\eta_{t,N^2}$. Therefore one can speculate that the LSD should be at least slowly suppressed. As an upper bound, the LSD will stay active at most until the stable stratification is so strong that $Rb_m<Rb_m^c$, which is when the SSD is shut down. However, determining the robustness of the LSD for intermediate stratification requires DNS.  

\section{Numerical Study and Results}\label{sec:Numerical}

\subsection{Numerical Setup}
\begin{figure}
    \centering
    \includegraphics[width=0.75\linewidth]{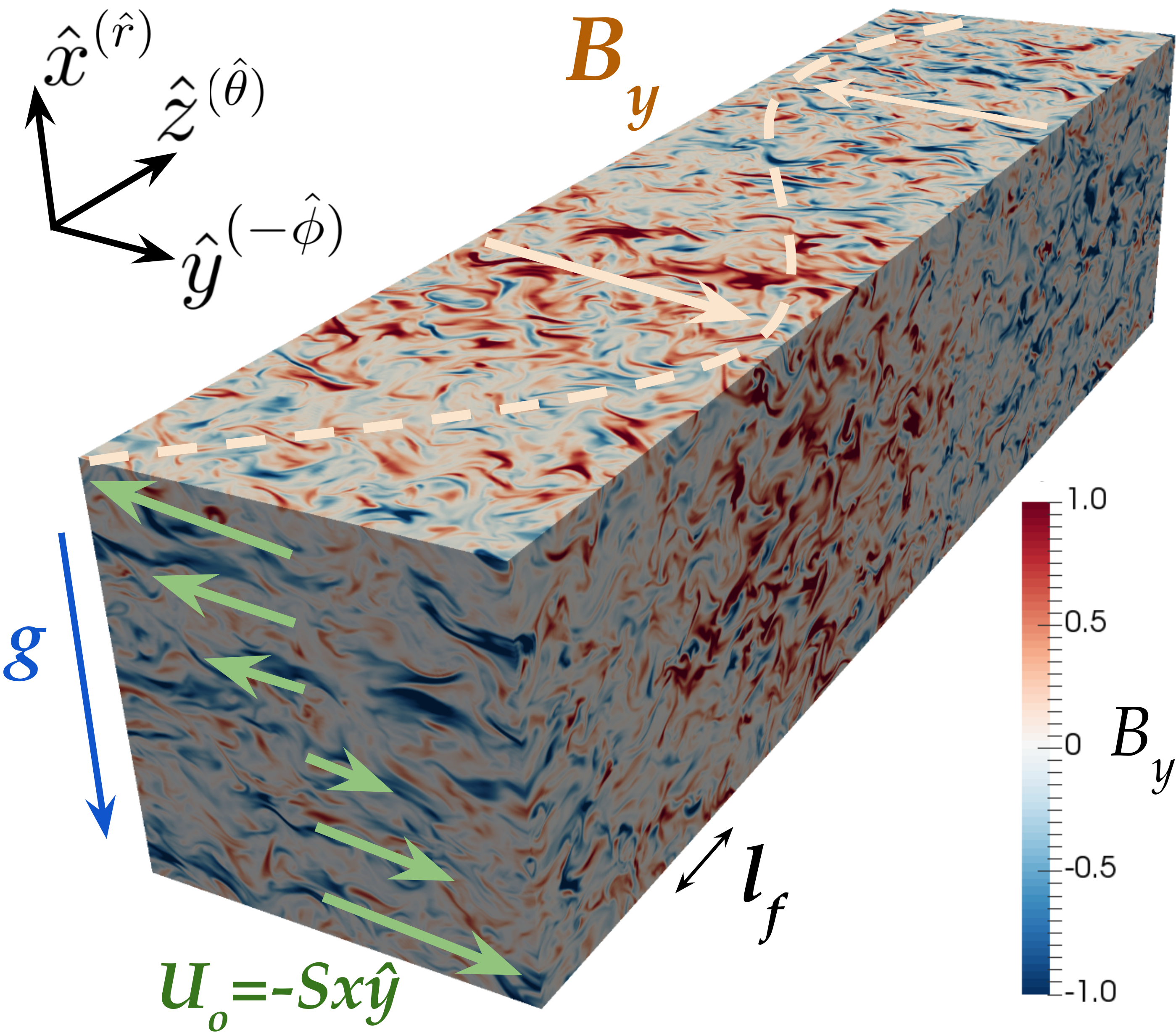}
    \caption{Annotated visualization of the y-component of the magnetic field $B_y(x,y,z)$ from a simulation with $Rm=210$, $Sh=1$, and $Fr^{-1}=3$. Boussinesq stratification (blue arrow) is imposed in the x-direction with Brunt-V$\ddot{\mathrm{a}}$is$\ddot{\mathrm{a}}$l$\ddot{\mathrm{a}}$ frequency $N$. A shear flow (green arrows) is imposed in the x-direction with profile $U_0=-Sx\hat{y}$. Forcing of the momentum equation at length scale $l_f=2\pi/k_f$ (black arrow) generates velocity fluctuations, which drive a SSD and generates magnetic fluctuations. The subsequent evolution of the large-scale velocity, $U(z)$, and magnetic field, $B(z)$ (light orange), is studied.}
    \label{fig:NumericalSetup}
\end{figure}

We use SNOOPY \citep{lesur2015snoopy}, a 3D pseudospectral code, with low-storage third-order Runge-Kutta time stepping and $3/2$ de-aliasing to carry out DNS of the MHD Boussinesq equations. Our default domain has a size $(L_x,L_y,L_z)=L(1,1,4)$.  Periodic boundary conditions are used in the $y$ and $z$ direction and shear periodic boundary conditions in the $x$ direction to model the imposed shear flow $U_0=-Sx\hat{y}$. The initial seed magnetic field is random at all scales and extremely weak ($E_b(t=0)=10^{-16}$) to allow self-consistent amplification by the SSD, if present. The momentum equation is driven with an isotropic, time-correlated forcing term $\sigma_f$ and the system is integrated in time (alternative forcing types give similar results). This is our model of a local patch of stably stratified turbulence in a differentially rotating stellar RZ, sketched in Figure \ref{fig:Diagram}. A visualization of a representative simulation is shown in Figure \ref{fig:NumericalSetup}.

The forcing is restricted to a waveband of width $\pi/L$ centered at $k_f=5\cdot 2\pi/L$, and has a correlation time $\tau_c=0.3$, chosen to satisfy the relation $u_{\mathrm{rms}}\approx 2\pi/(k_f\tau_c)$ at early times. The forcing wavenumber $k_f$ is chosen to allow more than an order of magnitude scale separation from the large scale at which the mean-fields are expected to grow (i.e. a scale separation of $k_fL_z/2\pi=20$), while still supporting a moderate turbulent cascade of the injected energy to the smallest, viscous scales. This turns out to be the minimum scale separation needed to capture the dominant LSD mode, as discussed in Section \ref{sec:IncoherentTest}. An explicit viscosity, resistivity, and thermal diffusivity (with $Pm=Pr=1$) is used to resolve the diffusive scales in the spectral code.

Signatures of the VD and LSD are most visible in two main diagnostics: 1) the isotropic energy spectra and 2) the time evolution of the energy in the large and small scales. The isotropic energy spectrum is defined in the standard way:
\begin{align}
    E_B(k,t)=\sum_{|\textbf{k}|\in[k-\frac{\pi}{L},k+\frac{\pi}{L}] }\frac{1}{2}|\hat{\textbf{B}}_\textbf{k}(t)|^2,
\end{align}
where $\hat{\textbf{B}}_\textbf{k}(t)$ is the Fourier transform of the magnetic field $\textbf{B}(\textbf{x},t)$ in the simulation. We define the wavenumber $k_s=2\pi/L$ as the separation between the large and small scales. Then, the large and small-scale magnetic energies are:

\begin{align}
    E_B(t)=\sum_{k\leq k_s}E_B(k,t), \quad   E_b(t)=\sum_{k> k_s}E_B(k,t).
\end{align}

The analogous definition holds for the kinetic energy spectra, $E_u(k,t)$, and the energy in the large and small-scale velocity fields $E_U(t)$ and $E_u(t)$, respectively. We additionally denote normalized magnetic energy with a tilde, e.g. $\tilde{E}_B(t)=E_B(t)/\overline{E_u(t)}$ where $\overline{E_u(t)}$ is the kinetic energy averaged over the last $50\tau_c$ of a simulation.
\begin{figure}
    \centering
    \includegraphics[width=\linewidth]{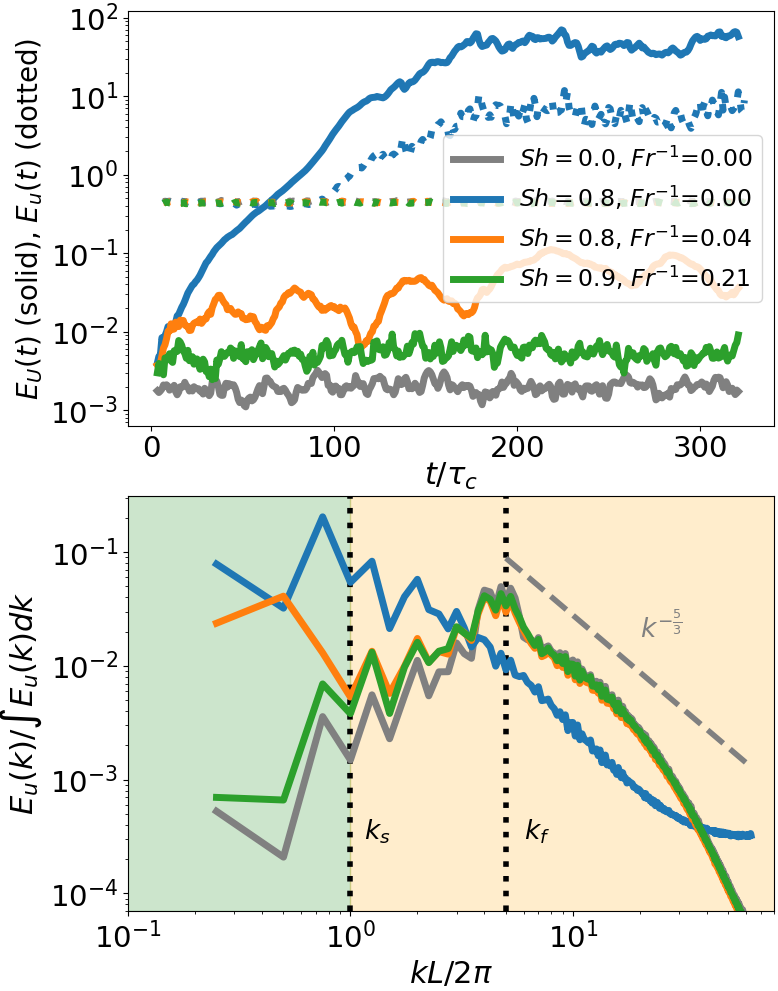}
    \caption{Examination of how the vorticity dynamo is easily stabilized with increasing stratification (increasing $Fr^{-1}$) along the direction of shear $\hat{x}$. Kinetic energy (top) and spectra (bottom) diagnostics of hydrodynamic simulations with varying shear $Sh$ and stratification $Fr^{-1}$ at fixed $Re\approx120$. Solid lines in the top panel are the energy in the mean field, $E_U$, while dotted lines are the energy in the small-scale velocity field, $E_u$. The spectra are computed from a snapshot at the last time point. Green and orange shaded regions in the bottom panel represent our definition of the large and small scales. The spectral resolution of the simulations is $N_x\times N_y\times N_z=192^2\times768$ modes.  }
    \label{fig:VD}
\end{figure}

\subsection{Hydrodynamic Vorticity Dynamo}

\subsubsection{Unstratified VD} We begin by confirming the results of \cite{Elperin2003} and \cite{PKapyla2009}  for the unstratified VD. We increase the shear parameter $Sh$ from $Sh=0$ to $Sh\approx1$ (by increasing $S=0$ to $S=4$) while keeping $Re\approx 120$ fixed and $Fr^{-1}=0$. These runs are hydrodynamic and the diagnostics are shown in gray and blue in Figure \ref{fig:VD}. For the energy evolution (top panel), the energy of the mean vorticity field $E_U(t)$ of the high $Sh\approx1$ run (solid blue) grows exponentially and saturates at orders of magnitude larger energies than the driven small-scale velocity turbulence, whose baseline level is $E_u(t)$ of the VD-stable runs (dotted gray, green, or orange). The normalized kinetic energy spectra at the last time point (bottom panel) also clearly reveals that the low wavenumber $k<k_s$ modes have more energy than the forcing wave numbers for the VD-unstable simulation (blue), while the same is not true for the run without shear (gray). Additionally, most of the energy in the large-scale modes is dominated by the y-component of the velocity field (not shown). These are the characteristic signatures of the VD that destabilizes this shearing box setup without stratification.

\subsubsection{Stably Stratified VD} To test the mean-field theory prediction, we slowly increase the strength of the background stable stratification and explore the effect on the hydrodynamic VD. We increase the stratification parameter $Fr^{-1}\in\{0.04,0.21\}$ while keeping $Sh\approx1$ and $Re\approx120$ fixed by increasing $N\in \{0.2,1\}$. Figure \ref{fig:VD} show the diagnostics in orange and green. Based on the evolution of $E_U(t)$ (top panel), the VD is already only marginally unstable at $Fr^{-1}=0.04$ (solid orange) and becomes fully stable at and above $Fr^{-1}=0.21$ (solid green). This is also seen in the kinetic energy spectra (bottom panel, orange and green) where the low wavenumber $k<k_s$ modes remain in subequiparition with the energy at the forcing wave number, except in the marginal case (orange) where they are modestly excited. Note that $Fr^{-1}\geq1$ corresponds to stratification being important at and above forcing scales (as well as a range of smaller scales), while $Fr^{-1}<1$ means stratification only affects scales larger than the forcing scale, which we call weak stratification here. Thus, our simulations qualitatively agree well with the mean-field theory prediction: weak stable stratification easily shuts down the VD. We note that adding magnetic fields to these simulations (not shown) does produce a SSD and LSD (only in the shearing cases), but their addition does not change the above results.

\subsection{Stably Stratified Large-scale Dynamo}\label{sec:NumericalMSC}
\subsubsection{Weak Stable Stratification} 
\begin{figure}
     \centering
    \includegraphics[width=\linewidth]{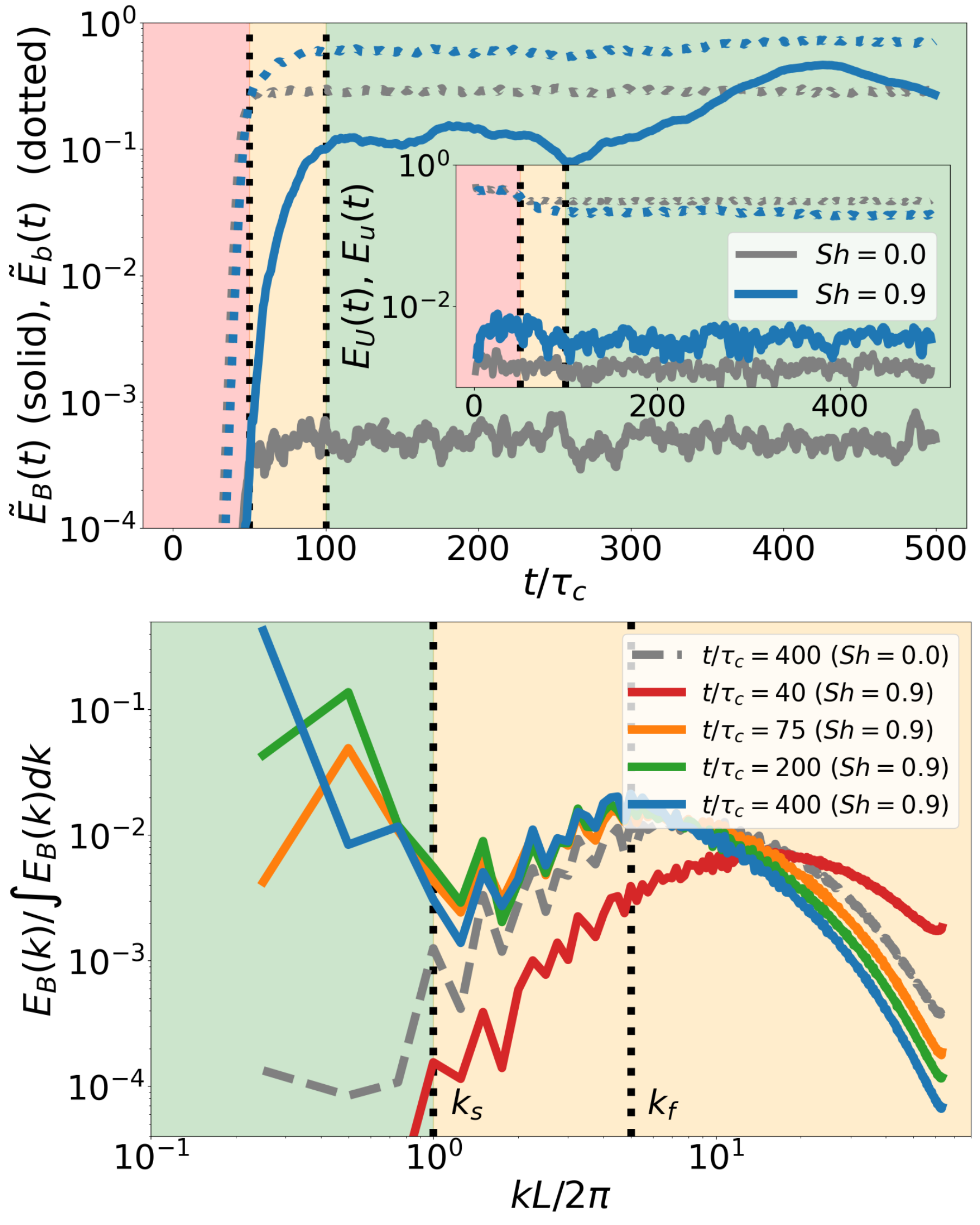}
    \caption{Comparing a simulation with shear ($Sh=0.9$) that is unstable to the LSD with a simulation with no shear ($Sh=0$). Both simulations have $Rm\approx120$, $Fr^{-1}=0.2$ with a spectral resolution of $N_x\times N_y\times N_z=192^2\times768$ modes. Top panel: solid lines are the energy in the mean field normalized by the kinetic energy, $\tilde{E}_B$, while dotted lines are the energy in the small-scale magnetic fields, $\tilde{E}_b$, similarly normalized. The inset plot shows the same diagnostics for the velocity field. Red, orange, and green shaded regions represent the SSD growth phase, LSD growth phase, and the LSD saturated phase. Bottom panel: normalized magnetic energy spectra at representative times (different colors). Green and orange shaded regions represent our definition of the large and small scales.  }
    \label{fig:SingleLSD}
\end{figure}

We turn to the case of the LSD in turbulence where the stratification is weak, but sufficient to stabilize the VD. The magnetic evolution of a VD-stable, weakly stably stratified system proceeds in three main phases as shown in the diagnostics of Figure \ref{fig:SingleLSD}. A simulation with $Rm\approx 120$, $Fr^{-1}=0.2$, and $Sh=0.9$ is compared against one with no shear ($Sh=0$) for reference. Both are stable to the VD as shown by the lack of energy growth in the mean velocity field $E_U(t)$ (solid lines of inset plot Figure \ref{fig:SingleLSD}). The first phase is the kinematic SSD phase (red shaded region, top panel), characterized by rapid exponential growth of $\tilde{E}_b(t)$ at early times (initially $\tilde{E}_b\approx 10^{-16}$), which ends and begins to saturate around $t/\tau_c\approx50$ for both cases. Note that in this regime the energy in the mean field $\tilde{E}_B$ also grows at the SSD growth rate due to the contribution of the infrared tail of the SSD eigenfunction in spectral space. After SSD saturation, the growth phase of the MSC effect begins (orange shaded region) and  $\tilde{E}_B(t)$ continues to grow at a slower LSD growth rate for the sheared case (solid blue curve, top panel), but completely stops growing for the no shear case (solid gray curve, top panel). 

\begin{figure}
    \centering
    \includegraphics[width=\linewidth]{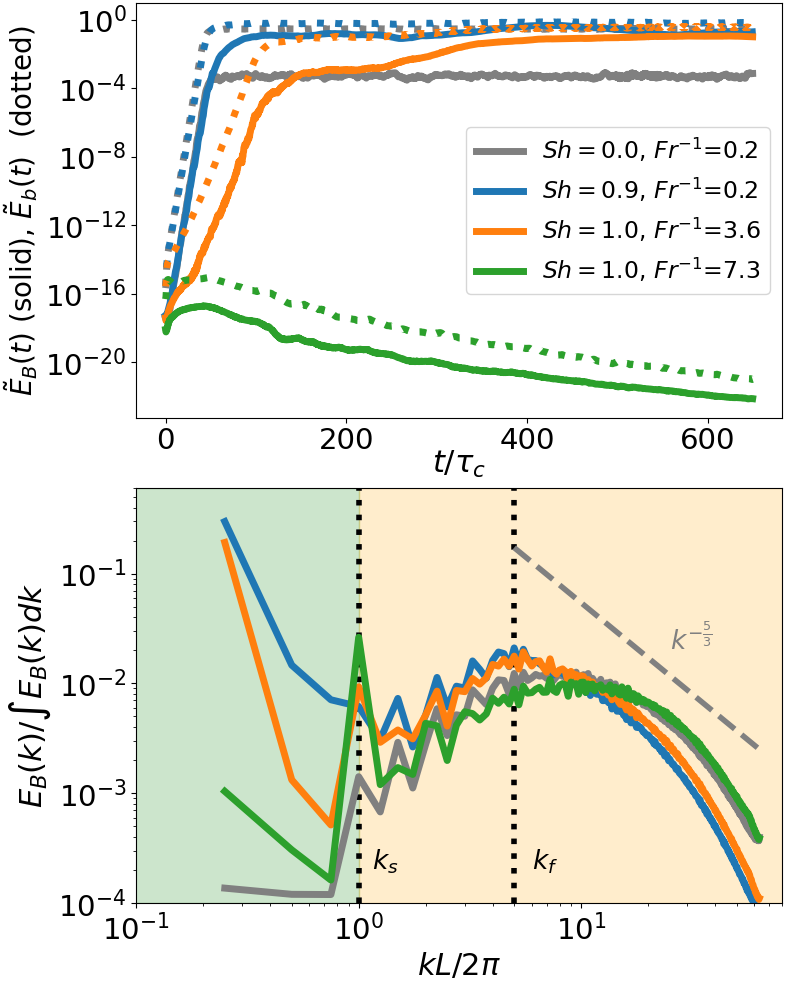}
    \caption{Exploring the effect of stable stratification on the LSD. The LSD appears robust as long as the SSD is unstable. Magnetic energy (top) and spectra (bottom) diagnostics of simulations with increasing stratification (increasing $Fr^{-1}$) at fixed $Rm\approx120$ and $Sh\approx0.9$. Solid lines in the top panel are the energy in the mean field, $\tilde{E}_B$, while dotted lines are the energy in the small-scale magnetic fields, $\tilde{E}_b$. The reference simulation with no shear ($Sh=0$) is shown in gray. All spectra are computed from a snapshot at the last time point. Green and orange shaded regions in the bottom panel represent our definition of the large and small scales. The spectral resolution of the simulations is $N_x\times N_y\times N_z=192^2\times768$ modes.  }
    \label{fig:MSC}
\end{figure}

We note that the LSD has a brief pseudo-linear phase from $\tilde{E}_B\approx5*10^{-4}$ to $\tilde{E}_B\approx 5*10^{-2}$ where the growth is quasi-exponential, but then transitions to a slower, non-linear growth phase as the LSD begins to saturate. DNS of LSDs driven by magnetic fluctuations can never have a long linear phase because the seed mean-fields inevitably start from moderate amplitudes set by the energy in the infrared wavenumbers of the saturated SSD spectrum. This is a domain-size-dependent effect---if the domain size was increased, the initial seed value of the mean field would decrease and lead to a longer linear phase. However, increasing the domain size any further is currently prohibitively expensive (but see Section \ref{sec:IncoherentTest} for a convergence test at a lower $Rm$).

The majority of the growth phase ends and the LSD saturation phase (green shaded region in top panel) begins for the sheared case around $t/\tau_c=100$ where the large-scale field energy undergoes a quasi-random behavior with slow oscillations on the timescale of hundreds of dynamical times (the origin of the quasi-random oscillations is not understood but is presumably related to the saturation mechanism). The difference compared to the no-shear case is striking as the LSD grows to be nearly in equipartition with the small-scale magnetic fields ($\tilde{E}_B\approx E_b$) around $t/\tau_c\approx 400$. In the no shear case, the random, large-scale fields remain several of orders of magnitude weaker than the small-scale fields ($\tilde{E}_B\ll \tilde{E}_b$) for all times. 
The difference is also clearly visible in the time evolution of magnetic energy spectra (bottom panel) where the energy in the lowest $k$ modes of the sheared case  steadily increases throughout the linear ( $t/\tau_c=75$) and saturation ( $t/\tau_c=200$) phases of the LSD and are nearly four orders of magnitude larger than that of the no shear case (dashed gray) at late times ($t/\tau_c=400$). Additionally, the lowest-$k$ modes of the sheared case are individually more than an order of magnitude larger in energy than the peak of the magnetic spectra at smaller scales near $k_f$ (this peak is simply that of the saturated SSD). These weakly stratified simulations are in good qualitative agreement with mean-field predictions in limit of perturbative stratification: the VD is easily suppressed while the LSD remains unstable. 

\subsubsection{Strong Stable Stratification} The next question to address is whether the LSD can operate in stable stratification that is non-perturbative and strong enough to affect the small-scale turbulence ($Fr^{-1}>1$). Figure \ref{fig:MSC} shows two revealing cases, one where the SSD is strongly suppressed but still active  ($Fr^{-1}\approx4$, orange) and another where the SSD has been shut down ($Fr^{-1}\approx7$, green). The case  where the SSD is shut down by strong stratification shows no LSD growth since both $\tilde{E}_b$ and $\tilde{E}_B$ decay, which confirms the expectation that without a source of magnetic fluctuations the MSC effect does not operate. The $Fr^{-1}\approx 4$ case however still shows a robust but slower LSD growth from $t/\tau_c\approx 200$ to $t/\tau_c\approx600$ when equipartition is reached $\tilde{E}_B\approx \tilde{E}_b$. The low $k$ modes of the magnetic spectra for the $Fr^{-1}\approx 4$ case (orange, bottom panel) are highly energized. The slower growth rate of the MSC effect is not surprising because the level of magnetic fluctuations is slightly lower due to the strong stratification, which can also been seen by the much lower growth rate of the SSD  of the $Fr^{-1}\approx 4$ case (dashed orange, top panel) compared to the e.g. $Fr^{-1}=0.2$ case (dashed blue, top panel). Nonetheless, the final saturation of the LSD is similar in both cases.
\begin{figure}
    \centering
    \includegraphics[width=\linewidth]{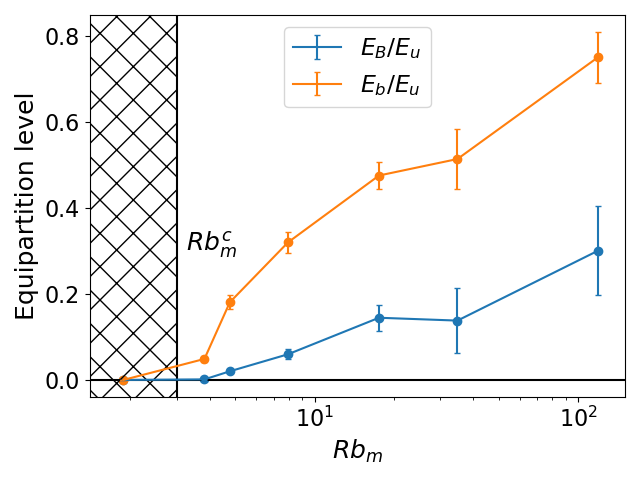}
    \caption{ The equipartition level of the saturated magnetic field with varying stratification. Colored lines show the energy fraction of large scale (blue) and small-scale (orange) magnetic fields relative to the turbulent kinetic energy as a function of $Rb_m=RmFr^2$. The simulations with increasing stratification (decreasing $Rb_m$) are carried out at constant $Rm\approx120$, $Sh\approx 1$. Energies and standard deviations are calculated long after saturation from the last $50\tau_c$ in each run. Hashed region denotes where the SSD is inactive due to strong stratification (for $Rb_m<Rb_m^c=3$ when $Pm=1$). }
    \label{fig:Equipartition}
\end{figure}

Suppression of the dynamo with increasing stratification is quantified in Figure \ref{fig:Equipartition} by showing the equiparition level of the large scale (blue, $E_B/E_u$) and small-scale  (orange, $E_b/E_u$) magnetic energy at saturation of the LSD (calculated at the end of each simulation) versus the stratification parameter $Rb_m=RmFr^2$. Figure \ref{fig:Equipartition} is generated from a series of simulations that vary $Fr$ at fixed $Rm\approx120$ and $Sh\approx1$. It appears that the LSD robustly operates in the non-perturbative limit of strong stratification ($Fr^{-1}>1$) with a near-equiparition saturation level of $E_B/E_u=O(10^{-1})$. The LSD begins to shut down when the SSD itself is strongly suppressed as $Rb_m$ approaches $Rb_m^c$ from the right. 

These idealized numerical results suggest that if a RZ with vertical shear contains stably stratified turbulence with a sufficiently large $Rb_m>Rb_m^c$ to sustain the SSD, the MSC should drive a large-scale magnetic field that reaches near-equiparition with the turbulent kinetic energy. 

\subsection{Role of the Incoherent Dynamo and Aspect Ratio}\label{sec:IncoherentTest}
\begin{figure}
    \centering
    \includegraphics[width=\linewidth]{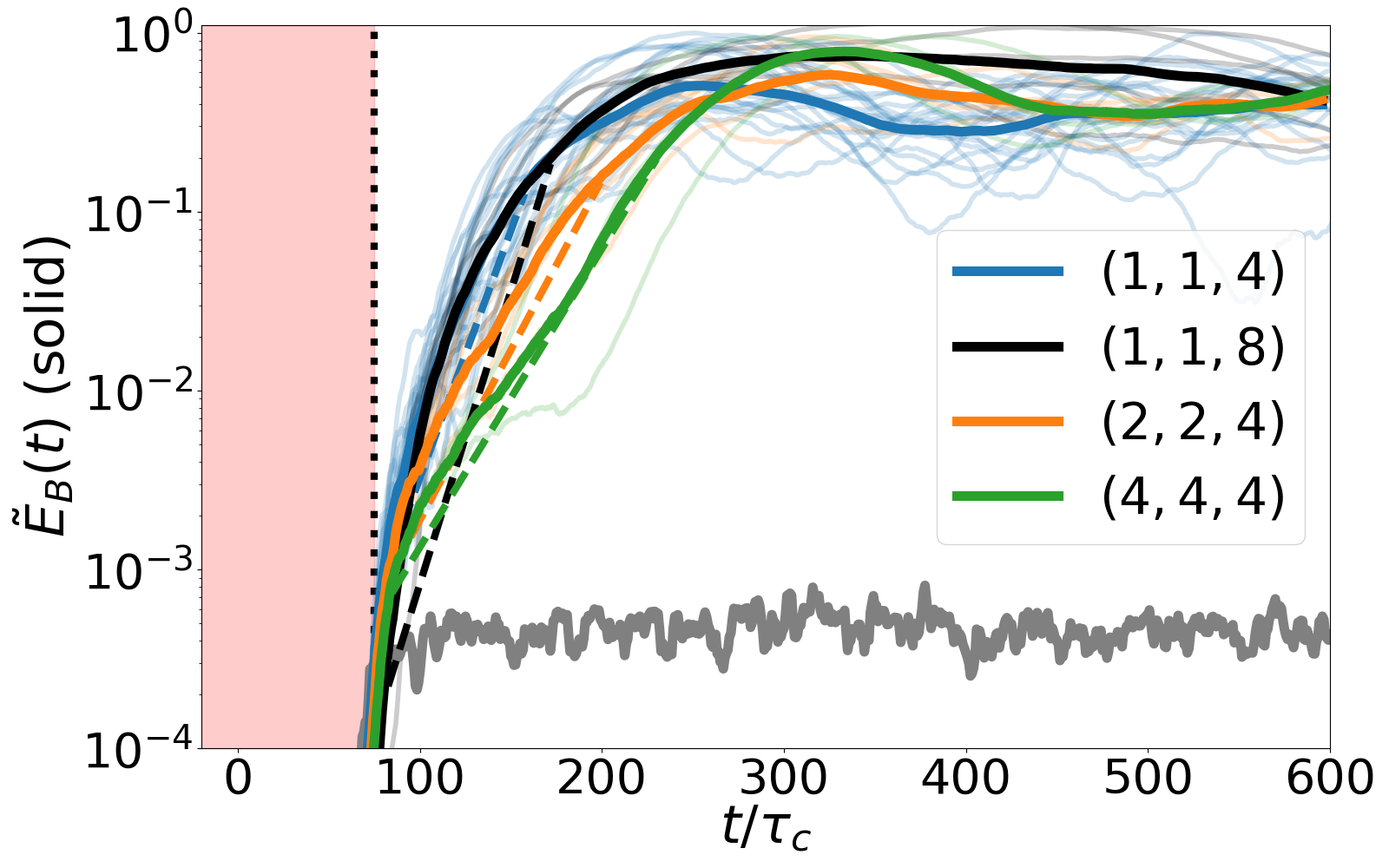}
    \includegraphics[width=0.95\linewidth]{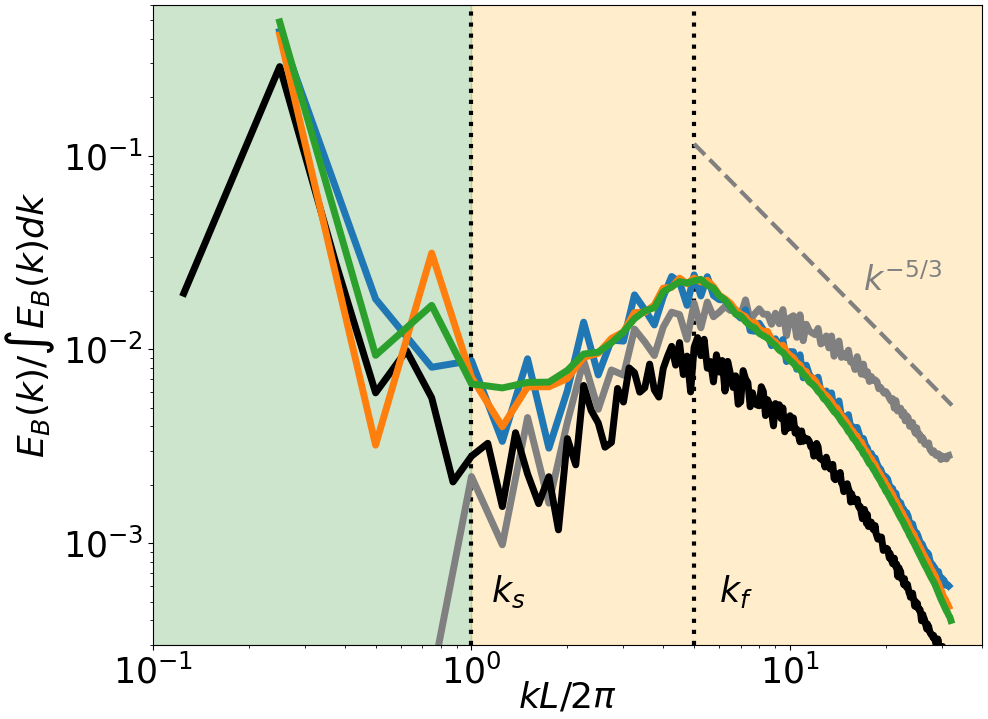}
    \caption{A validation test using ensembles of runs for different domain sizes ($L_x,L_y,L_z$) shows that incoherent dynamo effects are likely subdominant and that our fiducial domain size $(1,1,4)$ is sufficiently large to capture the dominant LSD mode. Top figure shows the evolution of the mean magnetic field energy for each case. Individual runs have low opacity while the run geometrically averaged over the ensemble is shown in full opacity for each aspect ratio. Dashed lines are the exponential fit to the LSD growth phase based on the average of the individually measured growth rates from the ensemble.  The red shaded region for $t/\tau_c\lesssim 75$ is the SSD growth phase. Bottom figure shows the magnetic energy spectra obtained at the end of the simulations.  All simulations have fixed values of $Rm\approx60$, $Sh\approx1.0$, and $Fr^{-1}\approx0.2$ and the spectral resolution is scaled with the aspect ratio (the fiducial $(1,1,4)$ simulation has $N_x\times N_y\times N_z=96^2\times384$ modes). Unlabeled gray curves are the no-shear ($Sh=0$) simulation for reference.  }
    \label{fig:Boxscan}
\end{figure}

\begin{figure}
    \centering
    \includegraphics[width=\linewidth]{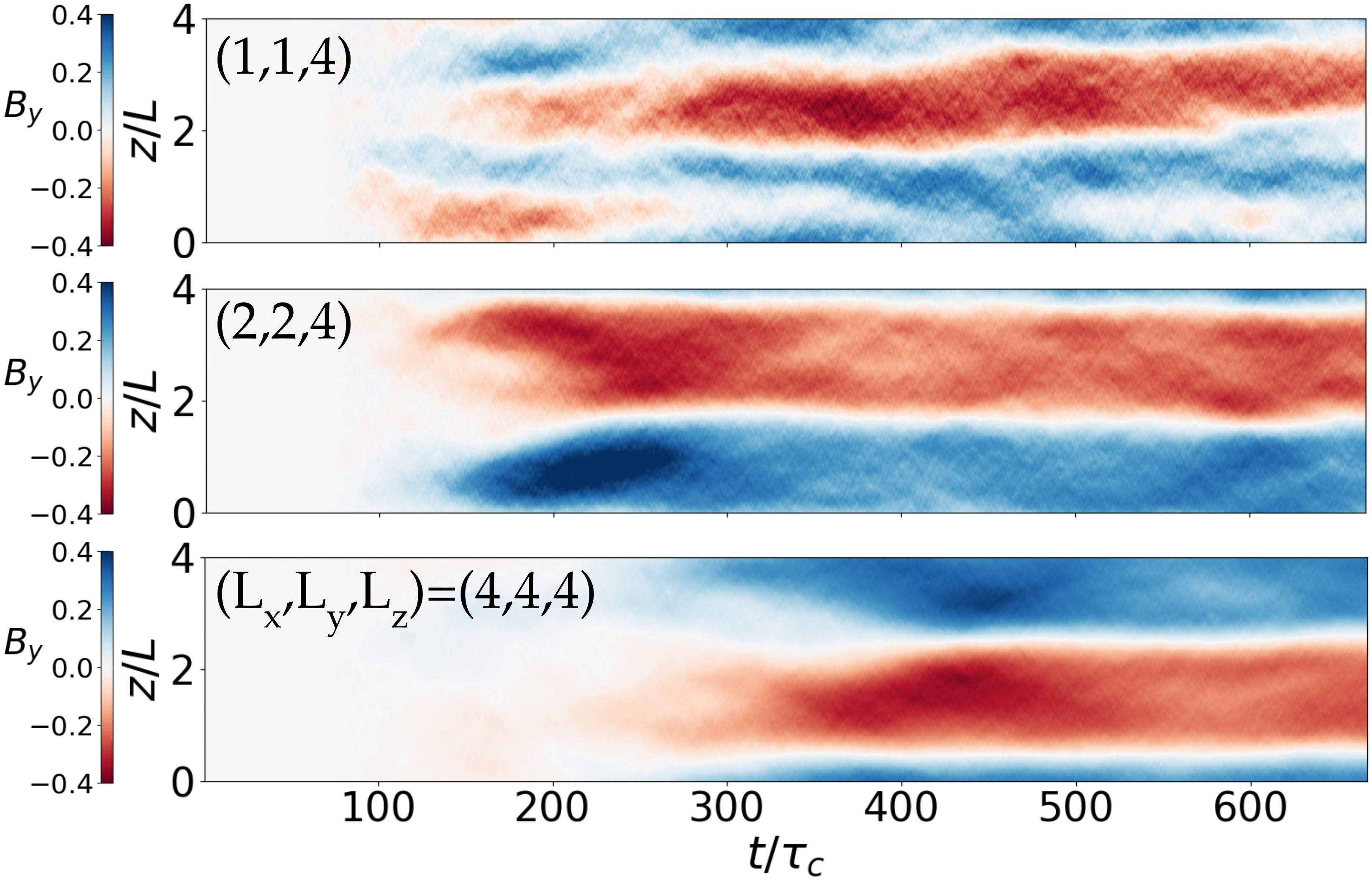}
    \caption{Visual validation that the phase of the saturated LSD mode remains coherent for long times. Time-space plots from single realizations of the y-component of the magnetic field $B_y(z,t)=\langle B_y(x,y,z,t)\rangle_{x,y}$ for three aspect ratios denoted in the top left corner of each plot. The $(1,1,4)$ run was carried out further to $t/\tau_c=1200$ with no qualitative change in behavior (not shown).  }
    \label{fig:SpacetimePlots}
\end{figure}

\subsubsection{Incoherent Dynamo}Finite size domains are susceptible to an additional LSD mechanism known as the incoherent dynamo and its contribution relative to a coherent shear-current effect remains a matter of debate. Studies that have attempted to directly measure $\eta_{yx}$ for different setups, parameter regimes, and test-field methods report different signs for $\eta_{yx}$ (see \citet{zhou2021} for a thorough review). \citet{singh2011transport} show that a kinematic shear-current effect cannot operate for any $Sh$ at low $Re\sim Rm\lesssim 1$, leaving only the $Rm\gg1$ regime as a possibility. Unfortunately, in this regime with full MHD background turbulence at moderate $Rm$, test-field methods that account for the magnetic background turbulence become only approximate and many variants are possible \citep{rheinhardt2010test,kapyla2022compressible}. Compressible MHD setups with driven turbulence, $Sh\lesssim 0.5$, $Re\gtrsim1$, $Rm\lesssim 100$, and scale separation $k_fL_z/2\pi\leq10$, find $\eta_{yx}$ to be slightly positive or slightly negative but not statistically different than zero in single realizations, concluding that incoherent effects are dominating  \citep{brandenburg2008magnetic,kapyla2022compressible}. Incompressible MHD simulations with driven turbulence at $Sh\lesssim 0.5$, $Re\sim Rm\lesssim 15$ and $k_fL_z/2\pi=6$ find a statistically significant negative value of $\eta_{yx}$ from ensembles of simulations \citep{Squire2016}, but the fitting assumptions used to reduce measurement noise have been debated. To add to the confusion, simulations of MRI driven MHD turbulence in a finite volume code \citep{shi2016saturation} and a moving mesh code \citep{zier2022simulating} have reported negative measurements of $\eta_{yx}$ while \citep{wissing2022magnetorotational} report a positive measurement using a smooth particle hydrodynamics code. \cite{zhou2021} resolve some of these discrepancies (at least at low and moderate $Re$) by showing that the kinetic contribution $\eta_{yx}^{(u)}$\footnote{The transport coefficient can be broken up into contributions from the velocity and magnetic fluctuations: $\eta_{yx}=\eta_{yx}^{(u)}+\eta_{yx}^{(b)}$} is sensitive to the spectral slope at low wavenumbers and becomes less efficient at counter-acting the generally negative $\eta_{yx}^{(b)}$ at higher $Re$ (i.e. $\eta_{yx}^{(u)}$ becomes less positive or even negative, possibly contributing to the MSC effect at high enough $Re$).

As an alternative approach in the $Re=Rm>50$ regime of this study, we do an indirect experiment that takes advantage of the volume independence/dependence of coherent/incoherent effects to test their relative contributions. The primary incoherent dynamo in a shear flow and non-helical turbulence is the stochastic-$\alpha$ effect, in which zero-mean fluctuations of the $\mathbf{\alpha}$ transport coefficients can drive growth of the variance of the mean-field $\langle \textbf{B}^2\rangle$ despite a zero ensemble average mean-field $\langle \textbf{B}\rangle=0$ \citep{vishniac1997incoherent,brandenburg2005astrophysical,heinemann2011large,Mitra2012}. These statistical properties make the stochastic-$\alpha$ effect dependent on the domain size unlike coherent dynamo effects \citep{squire2015coherent}.

The scaling of an incoherent dynamo with volume can be determined as follows. For the simplest case, consider zero mean $\langle \alpha_{yy}(t)\rangle=0$ fluctuations with a variance $\langle \alpha_{yy}(t)\alpha_{yy}(t')\rangle=D_{yy}\delta(t-t')$, which corresponds to a fluctuating EMF of the form $\mathcal{E}_y=\alpha_{yy}(t)B_y$ \citep{Mitra2012}. One can show that the fastest growing mode has a growth rate that scales as $\gamma^{\max}_\alpha\propto D_{yy}^{1/2}$ \citep{vishniac1997incoherent}. Because increasing the volume of the domain by a factor of $N$ decreases the variance $D_{yy}$ of the $\alpha_{yy}$ fluctuations by a factor of $N$ (assuming each of the $N$ sub-volumes are statistically independent), the stochastic dynamo growth rate must scale with the inverse square root of the domain volume $\gamma^{\max}_\alpha\propto V^{-1/2}$ \citep{squire2015coherent}. Therefore if our simulations are dominated by an incoherent effect, we should expect a significant decrease in the strength of the LSD when the volume is increased.

We carry out this experiment by progressively doubling or quadrupling the volume $V=L_xL_yL_z$ by changing the box aspect ratio $(L_x,L_y,L_z)/L=(1,1,4)$, $(1,1,8)$, $(2,2,4)$, and $(4,4,4)$ at fixed turbulent forcing scale and fiducial parameters $Rm\approx60$, $Sh\approx1.0$, $Fr^{-1}\approx0.2$. These parameters are stable to the VD, but unstable to the SSD and LSD. As discussed earlier in Section \ref{sec:NumericalMSC}, the growth rate of the LSD right after the SSD saturates is difficult to interpret because the pseudo-linear phase of the LSD is short in finite size simulations. To ameliorate this issue, we run an ensemble of simulations for each case to quantify the variance of the LSD evolution instead of relying on comparison between single realizations. The $(1,1,4)$, $(1,1,8)$, $(2,2,4)$, and $(4,4,4)$ cases each have $20$, $5$, $5$, and $3$ runs in their ensembles, respectively.

The results are shown in Figure \ref{fig:Boxscan} and the measured mean and standard deviation of the growth rate $\gamma_{\mathrm{LSD}}$ of each ensemble are reported in Table \ref{tab:LSDgrowthRates}. If the LSD was driven by an incoherent effect, we would expect the growth rate to decrease by a factor of $\sqrt{2}$, $2$, and $4$ for the $(1,1,8)$, $(2,2,4)$, and $(4,4,4)$ boxes compared to the $(1,1,4)$ box.  A strong volume dependence does not appear to be observed. The measured growth rates in Table \ref{tab:LSDgrowthRates} have a volume dependence $\gamma_{\mathrm{LSD}}\sim V^{-0.23\pm0.1}$ that is weaker than the $V^{-0.5}$ that would be theoretically expected if the observed LSD a purely an incoherent effect. The ensemble mean of the $(1,1,8)$ and $(2,2,4)$ cases in the top plot of Figure \ref{fig:Boxscan} both fall within the ensemble variance of the $(1,1,4)$ runs before beginning their slow, random oscillations in the saturated phase $t/\tau_c\gtrsim250$. The ensemble mean of the $(4,4,4)$ case has a noticeably slower quasi-exponential growth, but this may be due to a lack of statistical convergence since the ensemble is small with size of 3 (constrained by the increased computational cost of the larger domain). The ensemble mean appears to be heavily influenced by a single run whose growth stagnates for $125\lesssim t/\tau_c\lesssim200$ but then resumes growing at a comparable rate to the others for $200\lesssim t/\tau_c\lesssim250$. At late times, all aspect ratios saturate at similar energies and with similar magnetic spectra as shown in the bottom plot of Figure \ref{fig:Boxscan}.

\begin{table}
    \centering
    \begin{tabular}{c|c}
         Aspect Ratio & $\gamma_{\mathrm{LSD}}l_f/u_{\mathrm{rms}}$ \\
         \hline
         $(1,1,4)$ &  $(6.1\pm1.3)\cdot 10^{-2}$\\
         $(1,1,8)$ &  $(7.0\pm1.5)\cdot 10^{-2}$\\
         $(2,2,4)$ &  $(4.1\pm0.6)\cdot 10^{-2}$\\
         $(4,4,4)$ &  $(3.6\pm0.8)\cdot 10^{-2}$\\
    \end{tabular}
    \caption{Mean and standard deviations of the LSD growth rates $\gamma_{\mathrm{LSD}}$ for each  ensemble of simulations with different aspect ratios (whose time evolution is shown in Figure \ref{fig:Boxscan}). The growth rate for any single realization is measured between the time when the SSD saturates, $t_{\rm start}$, and when the mean field energy reaches $25\%$ of its maximum value $\widetilde{E}_B(t_{\rm end})=0.25\max (\widetilde{E}_B(t))$. We define SSD saturation by when the SSD growth rate falls below $10\%$ of its maximum value, $\gamma_{\rm SSD}(t_{\rm start})=0.1\max(\gamma_{\rm SSD}(t))$ where $\gamma_{\rm SSD}(t)=\partial_t \ln(E_b(t))$. Note that the size of the ensembles of the larger volume cases are smaller due to computational costs. }
    \label{tab:LSDgrowthRates}
\end{table}

Examining the phase variation of the mean field over time also offers an additional way to check for the presence of incoherence effects. An incoherent effect would generate a mean field with a randomly wandering phase \citep{squire2015coherent} while the dispersion relation of the coherent MSC effect predicts no phase variation since $\gamma^{\mathrm{MSC}}$ is purely real (Equation \ref{eq:MSCgammamax}). The space-time plots of individual runs in Figure \ref{fig:SpacetimePlots} demonstrate that the mean-field $B_y(z,t)$ maintains a constant phase for many LSD-growth time scales ($\gamma_{\rm LSD}^{-1}\sim 20\tau_c$ from Table \ref{tab:LSDgrowthRates}) in the saturation regime $t/\tau_c\gtrsim 300$ and that the mean-field of the larger volume runs appears progressively more coherent and smooth. The (1,1,4) and (2,2,4) domains are further run until $t/\tau_c=3000$ (see Appendix \ref{sec:LongRun}) for a thorough check on the long term behavior of the saturated LSD. Figure \ref{fig:LongRuns} shows that the (1,1,4) case begins to exhibit phase variation for $t/\tau_c>1000$, suggesting unknown behavior of the saturated LSD, the presence of an incoherent effect, or interaction of the two. However, the (2,2,4) exhibits no phase variation for the entire duration, which we interpret as the sufficiently large domain size where incoherent effects have become insignificant.

Overall, the weak volume dependence on the LSD growth rate and a fairly coherent constant-phase evolution with time suggests that incoherent effects have a subdominant contribution to the total LSD growth rate in our simulations. A contribution from an incoherent effect (likely strongest in the $(1,1,4)$ case) may explain the slight decrease in the LSD growth rate with increased simulation volume at a fixed forcing scale.  Large-scale dynamos in realistic astrophysical systems may be expected to be dominated by either incoherent or coherent effects depending on the scale separation, which is not always asymptotically large.

\subsubsection{Aspect Ratio and Convergence} While changing the aspect ratio, we can also check if our simulations are converged in the $z$-dimension \citep{yousef2008generation}. A comparison of the magnetic spectra in the bottom panel of Figure \ref{fig:Boxscan} of the $(1,1,4)$ and $(1,1,8)$ aspect ratios shows that the MSC effect is fully captured in our $(1,1,4)$ runs since the dominant mode is clearly the $k L/2\pi=1/4$ mode in the larger $(1,1,8)$ simulation (black). This qualitatively agrees with the dispersion relation Eq. \eqref{eq:MSCgammamax} which predicts a dominant wavenumber $k_{\max}$. Additionally, the time evolution of $E_B(t)$ of both cases is similar in the growth phase and after they enter the saturation regime $t/\tau_c\gtrsim250$ and slowly oscillate with similar amplitudes. In summary, we find that a minimum scale separation of $k_fL_z/2\pi=20$ is needed to capture the LSD at these parameters and that our fiducial $(1,1,4)$ simulations are sufficiently long in the $z$ direction. 

\subsection{Rm dependence of the MSC effect}

An outstanding problem in dynamo theory is understanding the amplitude and timescale of non-linear saturation of various LSD mechanisms in the limit of large $Rm$, as relevant to the astrophysical regime. LSDs based on the $\alpha$-effect face the well-known issue of catastrophic quenching, in which the amplitude and/or timescale of saturation scales strongly with the microscopic resistivity, suggesting an extremely weak LSD in the $Rm\gg1$ regime \citep{brandenburg2005astrophysical,rincon2019}. Although possibilities such as helicity fluxes through boundaries \citep{blackman2000constraints,vishniac2001magnetic,kleeorin2000helicity,brandenburg2002magnetichelicity}  and alternative scalings of small-scale helicity dissipation \citep{brandenburg2002magnetichelicity,brandenburg2005astrophysical,blackman2016magnetic} may resolve the problem, simulations so far have given mixed results \citep{brandenburg2005astrophysical,rincon2021helical}. Because the MSC effect is nonhelical, the usual helicity constraints that cause quenching do not apply (see Appendix \ref{sec:Helicity}). This means there is no a-priori reason that it should be catastrophically quenched, making it a promising mechanism that may operate at astrophysically large $Rm$. 
\begin{figure}
    \centering
    \includegraphics[width=\linewidth]{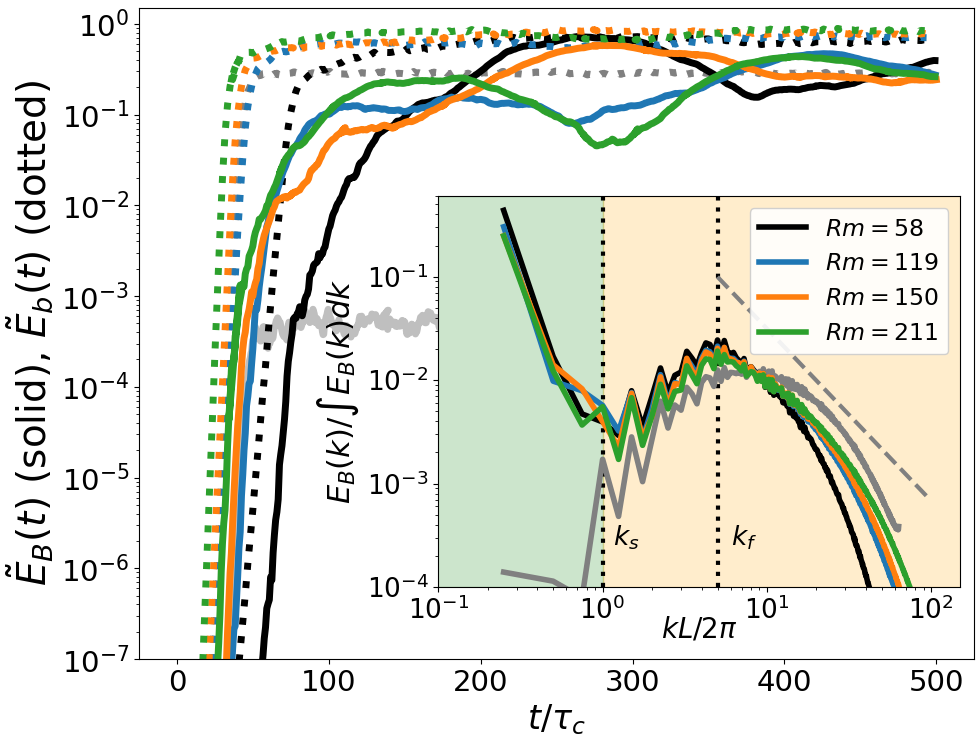}
    \caption{A test demonstrating that the MSC effect is likely free of quenching. Simulations with increasing $Rm$ (at $Pm=1$) are carried out by decreasing $\eta$ while keeping $Sh=0.9$ and $Fr^{-1}=0.2$ fixed. Magnetic energy (main figure) and spectra (inset figure) diagnostics are shown. The large scales of the magnetic spectra and saturation of the LSD appear to be unaffected. Unlabeled gray curves are the no-shear ($Sh=0$) simulation for reference. The spectral resolution of the $Rm=58,119$ runs is $N_x\times N_y\times N_z=192^2\times768$ modes and of the $Rm=150,211$ runs is $288^2\times1152$. }
    \label{fig:Quenching}
\end{figure}

To test the $Rm$ dependence of the MSC effect, we carry out a $Rm$ scan at $Pm=1$ to computationally accessible values, with all other parameters fixed. The fiducial parameters are  $Sh\approx1.0$ and  $Fr^{-1}\approx0.2$ as before. $Rm$ is increased by a factor of four from $Rm\approx50$ up to $Rm\approx200$ by decreasing $\eta$. A sign of catastrophic quenching in this test would be to observe the growth rate or saturation amplitude progressively decrease as a near power law with increasing $Rm$, as observed in helical dynamo simulations \citep{bhat2016unified,rincon2021helical}. Hence, the factor of four difference between the $Rm\approx50$ and $Rm\approx200$ simulations should have an easily discernible effect. The result is shown in Figure \ref{fig:Quenching} where we find that the LSD has no systematic $Rm$ dependence. Increasing $Rm$ only increases the SSD growth rate, as expected; which is seen through the progressively earlier saturation of $\tilde{E}_b(t)$ (dotted lines). This causes the LSD ($\tilde{E}_B(t)$, solid lines) to begin growth earlier, but they subsequently grow at a similar rate (albeit with the random oscillations discussed earlier) and each run saturates at a similar level. Indeed, these runs at different $Rm$ are not more dissimilar than runs from the ensemble of 20 simulations in Figure \ref{fig:Boxscan} at the same $Rm\approx 60$ for the $(1,1,4)$ case (after SSD saturation). Additionally, the magnetic spectra in the inset panel are independent of $Rm$ at large scales (low $k$). 

Thus our test finds no signs of catastrophic quenching with increasing $Rm$ up to the largest computationally accessible value of $Rm\approx200$. These numerical results, in combination with the above theoretical arguments, strongly suggest that the MSC effect is a dynamo mechanism free of catastrophic quenching that may operate in the astrophysicaly large $Rm$ regime applicable to RZs.

\section{Application to Radiative Zones}\label{sec:Application}
Our proposal is that the MSC effect may be an important non-linear dynamo mechanism that closes the global dynamo loop in differentially rotating RZs. Axisymmetric toroidal fields may grow from the shearing (from radial DR) of an axisymmetric poloidal if there is a source of non-helical magnetic fluctuations to drive the MSC effect and regenerate the axisymmetric poloidal field, closing the dynamo loop. We discuss two avenues for producing magnetic fluctuations (1) Tayler-instabilities of the toroidal field and (2) small-scale dynamo operating in stably stratified turbulence driven by horizontal shear instabilities of latitudinal DR. We assume that the MSC effect is agnostic to the instability that sources the magnetic fluctuations and that the  turbulence is predominantly non-helical at small scales. We leave studies of specific instabilities for future work.

\subsubsection{Tayler Instabilities}
In the framework of the TS dynamo, a sufficiently strong toroidal field is unstable to kink-type modes known as Tayler instabilities \citep{Tayler1973}, leading to magnetic turbulence. The Tayler-modes are non-axisymmetric and themselves cannot be directly sheared to regenerate the axisymmetric toroidal field, but instead are argued to contribute to a non-linear dynamo mechanism that regenerates the axisymmetric poloidal field \citep{zahn2007,fuller2019}. The MSC effect is a natural candidate mechanism because it is driven by magnetic fluctuations and is robust to stable stratification as shown in this study. Suggestions of an alpha based mechanism by previous studies \citep{zahn2007,fuller2019} are complicated by the known issue that small-scale magnetic fields generally suppress the alpha-effect and likely cause catastrophic quenching at the high $Rm$ regime relevant to RZs, as discussed earlier. The $Rm$-independent nature of the MSC effect makes it a promising alternative.

\subsubsection{Horizontal Shear Instabilities}
Another pathway way to generate magnetic fluctuations in a RZ is through the SSD operating in stably stratified turbulence driven by hydrodynamic instabilities. A likely possibility is horizontal shear instability of latitudinal DR. Vertical shear instabilities of radial DR, while generally stronger than latitudinal DR \citep{zahn1992circulation}, are most likely stabilized by the strong stratification in RZs \citep{garaudJtCS2021}. We propose that magnetic fluctuations from the SSD combined with radial DR may generate mean toroidal and poloidal fields through the MSC effect. Extrapolating from the results of Section \ref{sec:Numerical} suggests the toroidal field would reach near-equiparitition with the turbulence sourced by instability of the latitudinal DR. Note that our local shearing box setup cannot capture  such instabilities directly, because we do not include a horizontal shear, but heuristically captures the resulting small-scale turbulence through the external forcing term. Here, we use dimensionless numbers estimated based on helioseismology of the solar tachocline \citep{hughes2007solar}, the upper portion of the solar RZ, to examine the feasibility of our proposal.

The tachocline is approximately a thin spherical shell with radius $R=0.7R_\odot$, thickness $\Delta R\approx 10^{-2}R$, and differential rotation profile $\Omega(r,\theta)$. For the latitudinal DR, the differential angular velocity between the equator and the poles $(\Delta \Omega)_\theta$ is approximately $O(10^{-1})$ of the solar rotational frequency, i.e. $(\Delta \Omega)_\theta\approx 0.1\Omega_\odot$. For the radial DR, the differential angular velocity between the top and bottom of the tachocline at the equator is of similar strength $(\Delta \Omega)_r\approx 0.1\Omega_\odot$. Turbulence from the horizontal shear instabilities has an upper bound on the turbulent velocity $u_{\mathrm{rms}}\sim (\Delta \Omega)_{\theta}R$ and an effective forcing wavenumber likely comparable to $k_f\sim 2\pi/R$ \citep{cope2020dynamics,garaud2020horizontal}. The radial DR provides mean shear that is stable \citep{garaudJtCS2021} with a shear frequency that we estimate as $S=r(\partial\Omega/\partial r)\approx R(\Delta\Omega)_r/\Delta R$. The associated dimensionless numbers are:

\begin{equation}
    Re=8\cdot10^{13}\left(\frac{R}{5\cdot 10^8m}\right)^2\left(\frac{\Delta \Omega_{\theta}}{3\cdot10^{-7}\mathrm{ s}^{-1}}\right)\left(\frac{\nu}{10^{-3}m^2s^{-1}}\right)^{-1}
\end{equation}

\begin{equation}
    Fr=3\cdot 10^{-4}\left(\frac{\Delta \Omega_{\theta}}{3\cdot10^{-7}\mathrm{ s}^{-1}}\right)\left(\frac{N}{10^{-3}\mathrm{s}^{-1}}\right)^{-1}
\end{equation}

\begin{equation}
    Rb_m=7\cdot10^{4}\left(\frac{Pm}{10^{-2}}\right)\left(\frac{Re}{8\cdot10^{13}}\right)\left(\frac{Fr}{3\cdot 10^{-4}}\right)^{2}
\end{equation}
\begin{equation}
    Sh=10^2\left(\frac{\Delta\Omega_{r}/\Delta \Omega_\theta }{1}\right)\left(\frac{R/\Delta R}{10^{2}}\right)
\end{equation}
Comparing the  magnetic buoyancy Reynolds number $Rb_m=O(10^4)$ to the critical value $Rb_m^c=O(10)$ suggests that the SSD is unstable \citep{Skoutnev_2021}. The combination of a large shear number $Sh\gg1$ and a SSD providing magnetic fluctuations satisfies two the criteria for operation of the MSC effect. Therefore the MSC effect may generate near-equiparition magnetic fields with $B_{\phi}=O(1)T$, where we have use the equipartition estimate $B_{\phi}^2/2\mu_0\sim \rho u_{\mathrm{rms}}^2/2$ with $\rho\approx 200kg/m^3$. We expect these scalings to be reasonable in the interior of RZs of stars where latitudinal DR provides the dominant source of turbulence and the MSC effect is operating in isolation.

\subsubsection{Effect of Rotation and Low Prandtl Numbers}  
We briefly describe the possible modifications to our results by additional effects in RZs not explored in our study, that of rotation and low Prandtl numbers. These are likely subdominant to effects of strong radial shear and stable stratification in RZs. Rotation modifies the MSC effect directly through an orientation-dependent contribution to $\eta_{yx}$ and will have a significant effect if $S/\Omega\lesssim 1$ \citep{squire2015coherent}. While this is likely not important in the tachocline where $S/\Omega\approx ((\Delta\Omega)_r/\Omega_\odot) (R/\Delta R)\approx 10 $, it may be important in other stars with weaker radial DR or faster rotation. Another important source of uncertainty are the effects of low Prandtl numbers typical of stellar interiors ($Pm=O(10^{-2})$, $Pr=O(10^{-6})$ in the tachocline \citep{garaudJtCS2021}). A lower $Pm$ generally makes a SSD more difficult to sustain \citep{Iskakov_lowpm}, which is captured by the dependence of $Rb_m^c(Pm)$ on $Pm$ \citep{Skoutnev_2021}. However, for the LSD, calculations with the SOCA have found that the MSC effect is not sensitive to $Pm$ \citep{squire2015electromotive}. Unfortunately, confirming these results for the LSD with DNS at low $Pm$ is currently impractical due to an even larger requirement for $Re$. On the other hand, a low $Pr=\nu/\kappa$ is expected to make the SSD and therefore the LSD more unstable \citep{skoutnev2022critical}. A higher thermal diffusivity, $\kappa$, decreases the effect of stratification and leads to more isotropic turbulence, which generally enables more efficient dynamo action.

\section{Conclusion}\label{sec:Conclusion}
We examine the effects of stable stratification on mean-field dynamos with a particular focus on the magnetic shear-current effect and our results suggest that it can likely operate in the differentially rotating and stably stratified plasma of a stellar radiative zone. The dynamo loop closed by the magnetic shear-current effect \citep{rogachevskii2004nonlinear,squire2015generation,squire2015electromotive,Squire2016} generates toroidal field from the shearing of a poloidal field and regenerates poloidal field from the toroidal field through statistical correlations of local, non-helical MHD turbulence. Our analysis is based on idealised theory and simulations modeling a local section of a RZ, providing evidence to support a broader picture of dynamos in RZs. The key pieces of evidence are: 

\begin{enumerate}
\item Perturbative mean-field dynamo theory, when extended to include stable stratification (along the direction of shear), predicts the MSC instability remains robust with a decreased growth rate compared to the unstratified case. The hydrodynamic vorticity dynamo, however, is rapidly stabilized by stratification.
\item Shearing box simulations show that a mean shear combined with an unstable small-scale dynamo in stably stratified turbulence is unstable to a LSD, qualitatively agreeing with mean-field dynamo theory. The simulations also confirm that the hydrodynamic vorticity dynamo is stabilized by the addition of weak stable stratification.
\item Simulations show that the energy in the mean (toroidal) magnetic field at saturation is comparable to the turbulent kinetic energy.
\item A numerical scan of the magnetic Reynolds number demonstrates that the LSD does not suffer catastrophic quenching. In particular, the saturation time and amplitude are found to be independent of $Rm$. This is expected because there are no obvious constraints arising from magnetic helicity conservation on a non-helical dynamo mechanism.
\end{enumerate}

Put together, these idealized results suggest that the MSC effect in a RZ requires (1) a source of non-helical magnetic fluctuations and (2) sufficient radial differential rotation (velocity shear). The resulting mean-field should saturate at near-equipartition with the magnetic fluctuations in a manner that is free from catastrophic quenching (a significant issue for helicity-based alpha dynamo mechanisms).

Extrapolating our results from a local shearing box model to a realistic RZ, we propose two pathways to provide non-helical magnetic fluctuations for operation of a large-scale dynamo through the MSC effect in a region of radial DR. The first is Tayler instabilities \citep{Tayler1973,markey1973adiabatic} of the toroidal field, which directly result in the magnetic fluctuations that we propose may drive the MSC effect and regenerate axisymmetric poloidal field, thereby closing the Tayler-Spruit dynamo \citep{spruit2002,fuller2019}. The second is the small-scale dynamo operating in stably stratified turbulence, which may be driven by instabilities of latitudinal DR \citep{zahn1974rotational,zahn1992circulation,prat2013turbulent,prat2014shear,cope2020dynamics,garaud2020horizontal}. The reality may be a mixture of the two processes, and perhaps others. Although significant uncertainties remain (effects of spherical geometry, the helicity fraction of instability-driven magnetic turbulence, low Prandtl numbers etc.), the near-equiparition saturation, robustness to stable stratification and immunity to catastrophic quenching of the MSC effect make it worthy of further consideration in more complex global dynamo models in stellar RZs.

\section{Data Availability}
All numerical data was generated using the publicly available SNOOPY code \citep{lesur2015snoopy}. The data underlying this article will be shared on reasonable request to the corresponding author.
\acknowledgments
A.B. was supported by the DOE Grant for the Max Planck Princeton Center (MPPC). J.S. was supported by a Rutherford Discovery Fellowship RDF-U001804, which is managed through the Royal Society Te Ap\=arangi. V.S. was supported by Max-Planck/Princeton Center for Plasma Physics (NSF grant PHY-1804048). We thank Axel Brandenburg and members of the "Magnetic Field Evolution in Low Density or Strongly Stratified Plasmas" conference for helpful discussions and Kailey Whitman for help with generation of the diagrams.

\appendix

\section{Electromotive force calculation for the MHD Boussinesq equations}\label{sec:SOCA}
The general setup for calculating mean field transport coefficients using the second order correlation approximation (SOCA) supposes a bath of homogeneous and isotropic velocity and magnetic field fluctuations in the presence of anisotropic perturbations such as shear flows $\textbf{U}_0$ of form $U_i=U_{ij}x_j$ (with no vertical component $U_x=0$ in the Boussinesq case), rotation ($\mathbf{\Omega}$), and stable stratification (in the $\hat{x}$ direction with Brunt-Vaisala frequency $N$). While rotation is not included in the main paper, it is simple to include it here to demonstrate the mean-field MHD Boussinesq framework for the full problem. We assume the mean velocity field does not evolve and write the total fields as $\textbf{U}_T=\textbf{U}_0+\textbf{u}$, $\textbf{B}_T=\textbf{B}+\textbf{b}$, and $\Theta_T=\theta$ ($\Theta=0$ because there cannot be any mean vertical flows that could drive $\Theta$). The mean field induction equation then is given by: 
\begin{equation}
\partial_t {\textbf{B}} =\nabla \times({\textbf{U}_0}\times {\textbf{B}})+\nabla \times\mathcal{E} +\eta\nabla^2 {\textbf{B}},
\end{equation}
where the EMF is: 
\begin{equation}\label{eq:EMF}
    \mathcal{E}=\langle \textbf{u}\times \textbf{b}\rangle=\mathcal{E}(\textbf{B}).
\end{equation}
The fluctuation equations for $\textbf{u}$, $\textbf{b}$, and $\mathbf{\theta}$ are obtained by subtracting the mean field equations from those of the total fields:

\begin{align}\label{eq:flucts1}
\partial_t \textbf{u}+&\textbf{u}\cdot\nabla \textbf{U}_0+\textbf{U}_0\cdot\nabla \textbf{u}+(\textbf{u}\cdot\nabla \textbf{u})'
+2\mathbf{\Omega}\times \textbf{u}=-\nabla p+\theta\hat{x}\\ \nonumber
&+\textbf{B}\cdot\nabla \textbf{b}+\textbf{b}\cdot\nabla \textbf{B}+(\textbf{b}\cdot\nabla \textbf{b})'+\nu\nabla^2\textbf{u}+\sigma_f,\nonumber
\end{align}

\begin{align}\label{eq:flucts2}
\partial_t \textbf{b}=&\nabla \times (\textbf{U}_0\times \textbf{b}+\textbf{u}\times \textbf{B}+(\textbf{u}\times \textbf{b})')+\eta\nabla^2\textbf{b},
\end{align}

\begin{align}\label{eq:flucts3}
\partial_t \theta+&\textbf{U}_0\cdot\nabla \theta+(\textbf{u}\cdot\nabla \theta)'
=-N^2u_x+\kappa\nabla^2\theta,
\end{align}

\begin{equation}\label{eq:flucts4}
\nabla \cdot \textbf{u}=0, \; \nabla \cdot \textbf{b}=0,
\end{equation}
where we have used the notation $(A)'=A-\langle A \rangle$.

Following \citet{radler2006mean}, the EMF $\mathcal{E}$ can be Taylor expanded and linearly related to the mean field $B_i$ and its derivative $B_{i,j}$ (i.e. $\mathcal{E}_i=a_{i,j}B_j+b_{ijk}B_{j,k}...$) assuming sufficient scale separation. This then provides a closure of the mean field induction equation. Taking into account all symmetry properties, the most general form of transport coefficients linearly relating $\mathcal{E}$ and $\textbf{B}$ is:
\begin{align}
\mathcal{E}=&-\alpha^{(0)}_H\mathbf{B}-\alpha^{(D)}_H D_{ij}B_j-\gamma^{(\Omega)}_H \mathbf{\Omega} \times \mathbf{B}-\gamma^{(W)}_H \mathbf{W}\times \mathbf{B}\nonumber\\
&-\alpha^{(\Omega)}_1(\hat{\mathbf{g}}\cdot \mathbf{\Omega})\mathbf{B}-\alpha^{(\Omega)}_2[(\hat{\mathbf{g}}\cdot \textbf{B})\mathbf{\Omega}+(\mathbf{B}\cdot \mathbf{\Omega})\hat{\mathbf{g}}]-\alpha^{(W)}_1(\mathbf{g}\cdot \mathbf{W})\mathbf{B} \nonumber\\
&-\alpha^{(W)}_2[(\hat{\mathbf{g}}\cdot B)\mathbf{W}+(\mathbf{B}\cdot \mathbf{W})\hat{\mathbf{g}}]\nonumber\\
&-\alpha^{(D)}(\epsilon_{ilm}D_{lj}\hat{g}_m+\epsilon_{jlm}D_{li}\hat{g}_m)B_j\nonumber\\
&-(\mathbf{\gamma}^{(0)}+\gamma^{(\Omega)}\hat{\mathbf{g}}\times \mathbf{\Omega}+\gamma^{W}\hat{\mathbf{g}}\times \mathbf{W}+\gamma^{(D)}D_{ij}\hat{g}_j)\times \textbf{B} \nonumber \\
&-\beta^{(0)}\mathbf{J}-\beta^{(D)}D_{ij}J_j-(\delta^{(W)}\textbf{W}+\delta^{(\Omega)}\mathbf{\Omega})\times \textbf{J}\nonumber\\
&-(\kappa^{(W)}\textbf{W}+\kappa^{\Omega}\mathbf{\Omega})_j(\mathbf{\nabla} \textbf{B})^{(s)}_{ji}-2\kappa^{(D)}\epsilon_{ij k}D_{kr}(\mathbf{\nabla} \textbf{B})^{(s)}_{jr}.
\end{align}

The calculation of $\mathcal{E}$ at this point can be carried out with the help of the SOCA. The SOCA assumes that the higher order correlation terms (the primed terms such as $(\textbf{u}\cdot\nabla \textbf{u})'$) are small compared to terms involving the mean fields (e.g. $\textbf{u}\cdot \nabla \textbf{U}_0$) and can be neglected. This will lead to linear time evolution equations for $\textbf{u}$, $\textbf{b}$, and $\mathbf{\theta}$ that can be solved perturbatively. The perturbation expansion is done on both $\textbf{u}$ and $\textbf{b}$ around their background, homogeneous fluctuation values $\textbf{u}_0$ and $\textbf{b}_0$ and then substituted into Equation \eqref{eq:EMF} for the EMF. While statistics of the background fluctuations will be assumed, they physically arise from the forcing term $\sigma_f$ of the momentum equation (sustaining $\textbf{u}_0$) and a resulting small-scale dynamo (sustaining $\textbf{b}_0$ in equiparition). The forcing itself models some hydrodynamic instability, such as horizontal shear instabilities discussed in the main article. In general, the background fluctuations have both helical and non-helical components, which we include in the calculation. To allow the perturbative expansion, all anisotropic parameters such as $N^2,S,\Omega$ are considered to be small. With the notation $\textbf{u}=\textbf{u}_0+\textbf{u}^{(0)}+\textbf{u}^{(1)}...$ for the expansion of $\textbf{u}$ (as well as $\textbf{b}$ and $\theta$), $\mathcal{E}$ to second order is:

\begin{equation}\label{EMF}
    \mathcal{E}=\langle \textbf{u}_0\times \textbf{b}^{(0)}\rangle+\langle \textbf{u}^{(0)}\times \textbf{b}_0\rangle+\langle \textbf{u}^{(0)}\times \textbf{b}^{(0)}\rangle+\langle \textbf{u}_0\times \textbf{b}^{(1)}\rangle+\langle \textbf{u}^{(1)}\times \textbf{b}_0\rangle,
\end{equation}
where $\langle \textbf{u}_0\times \textbf{b}_0\rangle=0$ is assumed.

The calculation of $\mathcal{E}$ is carried out in Fourier space and explained in detail in \citet{radler2006mean}. We only give a brief description of the approach in order to point out how we handle the new addition of the buoyancy equation and the bouyancy term in the momentum equation in the SOCA formalism. To begin, we write out the evolution equations in real space for each order by applying the expansion to the fluctuation equations and then using the SOCA where applicable. The background, zeroth order, and first order equations are shown below.

\subsection{Background Turbulence}
The homogeneous, background fluctuations satisfy:
\begin{equation}
\partial_t \textbf{u}_0+(\textbf{u}_0\cdot \nabla \textbf{u}_0)^{'} =-\nabla p_0 +\theta_0\hat{x}+(\textbf{b}_0\cdot\nabla \textbf{b}_0)^{'}+\nu\nabla^2 \textbf{u}_0+\sigma_f,
\end{equation}

\begin{equation}
\partial_t \textbf{b}_0 =\nabla \times(\textbf{u}_0\times \textbf{b}_0)'+\eta\nabla^2 \textbf{b}_0,
\end{equation}

\begin{equation}
\partial_t \theta_0+\textbf{u}_0\cdot \nabla \theta_0=\kappa\nabla^2\theta_0.
\end{equation}
While the homogeneous velocity and magnetic fluctuations $\textbf{u}_0$ and $\textbf{b}_0$ are assumed to be in a steady state driven by the forcing $\sigma_f$ and a SSD, the buoyancy equation for the buoyancy fluctuations does not have any source term. Therefore, any initial buoyancy variable fluctuations $\theta_0(t=0)$ will be passively advected and thermally diffuse to zero after a transient phase. These background equations themselves are not used in the calculation for $\mathcal{E}$, but a model for the homogeneous statistics of the background turbulence will be used. 
\subsection{Zeroth Order}

The SOCA approximation assumes terms such as $(\textbf{u}^{(0)}\cdot \nabla \textbf{u}^{(0)})^{'}$ are much smaller than $\textbf{u}_0\cdot \nabla \textbf{U}_0$ so then the equations for the zeroth order fluctuations become:

\begin{align}\label{momEqZero}
(\partial_t-\nu\nabla^2) \textbf{u}^{(0)}=&-(\textbf{U}_0\cdot \nabla \textbf{u}_0+\textbf{u}_0\cdot \nabla \textbf{U}_0) -\nabla p^{(0)}-2\mathbf{\Omega}\times \textbf{u}_0 \nonumber \\&+ \theta^{(0)}\hat{x}+\textbf{B}\cdot\nabla \textbf{b}_0+\textbf{b}_0\cdot\nabla \textbf{B},
\end{align}

\begin{equation}
(\partial_t-\eta\nabla^2) \textbf{b}^{(0)} =\nabla \times(\textbf{U}_0\times \textbf{b}_0+\textbf{u}_0\times \textbf{B}),
\end{equation}

\begin{equation}\label{bouyEqZero}
(\partial_t-\kappa\nabla^2) \theta^{(0)}=-N^2 u_{0,x},
\end{equation}
where we have dropped the $\textbf{U}_0\cdot \nabla \theta_0$ term in the buoyancy Equation \ref{bouyEqZero} because the $\theta_0$ fluctuations are zero after a transient phase as discussed above. 
\subsection{First Order} Similarly, the equations for the first order fluctuations are:

\begin{align}\label{momEqFirst}
(\partial_t-\nu\nabla^2) \textbf{u}^{(1)}=&-(\textbf{U}_0\cdot \nabla \textbf{u}^{(0)}+\textbf{u}^{(0)}\cdot \nabla \textbf{U}_0) -\nabla p^{(1)}-2\mathbf{\Omega}\times \textbf{u}^{(0)}\nonumber\\& + \theta^{(1)}\hat{x}+\textbf{B}\cdot\nabla \textbf{b}^{(0)}+\textbf{b}^{(0)}\cdot\nabla \textbf{B},
\end{align}

\begin{equation}
(\partial_t-\eta\nabla^2) \textbf{b}^{(1)} =\nabla \times(\textbf{U}_0\times \textbf{b}^{(0)}+\textbf{u}^{(0)}\times \textbf{B}),
\end{equation}

\begin{equation}\label{bouyEqFirst}
(\partial_t-\kappa\nabla^2) \theta^{(1)}=-\textbf{U}_0\cdot \nabla \theta^{(0)}-N^2u_x^{(0)}.
\end{equation}

\subsection{Calculation of $\mathcal{E}$}

The calculation of $\mathcal{E}$ proceeds exactly as in \citet{squire2015electromotive} except with the addition of the buoyancy terms. The zeroth and first order fluctuation equations \ref{momEqZero} through \ref{bouyEqFirst} are transformed to Fourier space and substituted into the Fourier space version of EMF equation \ref{EMF} (see \citet{squire2015electromotive} and \citet{radler2006mean} for extensive details). Because $\theta_0$ fluctuations diffuse away and are not relevant for the background turbulence, it is unnecessary to make an additional model of the turbulent statistics between $\theta_0$ and either $\textbf{u}_0$ or $\textbf{b}_0$. The lack of $\theta_0$ is what allows $\theta^{(0)}$ and $\theta^{(1)}$ to be solved for in terms of $\textbf{u}_0$ and $\textbf{u}^{(0)}$ using the buoyancy equations \ref{bouyEqZero} and \ref{bouyEqFirst} and then substituted directly into the momentum equations \ref{momEqZero} and \ref{momEqFirst}, respectively. The net result of weak stably stratification therefore is added terms proportional to $N^2$ in the momentum equations \ref{momEqZero} and \ref{momEqFirst}.

We use the open source VEST package \citep{squire2014vest} for Mathematica to carry out the calculation, similar to \citet{squire2015electromotive}, including both helical and non-helical portions of the background velocity and magnetic fluctuations. We report only the four transport coefficients that we find are modified by stable stratification, the isotropic turbulent resistivity and alpha coefficient. All other transport coefficients are identical to the result in \citet{squire2015electromotive}. The modified coefficients in Fourier space are shown below: 

\begin{align}
    (\tilde{\beta}^{(0)})_u&=\frac{u_{\mathrm{rms}}l_f}{\eta}\Bigg[\frac{\tilde{k}^2}{3(\tilde{k}^4+q^2\tilde{\omega}^2)}\\&+\frac{3(N\tau_c)^2\tilde{k}^2(\tilde{k}^2\frac{Pm^2}{Pr}-q^2\tilde{\omega}^2)}{10(\tilde{k}^4+q^2\tilde{\omega}^2)(\tilde{k}^4Pm^2+q^2\tilde{\omega}^2)(\tilde{k}^4\frac{Pm^2}{Pr^2}+q^2\tilde{\omega}^2)}\Bigg],\nonumber
\end{align}
\begin{align}
    (\tilde{\beta}^{(0)})_b=(N\tau_c)^2\left[\frac{q^2\tilde{k}^2(\tilde{k}^4\frac{Pm^3}{Pr^2}-(\frac{Pm}{Pr}+2Pm)q^2\tilde{\omega}^2)}{60(\tilde{k}^4Pm^2+q^2\tilde{\omega}^2)^2(\tilde{k}^4\frac{Pm^2}{Pr^2}+q^2\tilde{\omega}^2)}\right],
\end{align}
\begin{align}
    (\tilde{\alpha}^{(0)}_H)_u&=\frac{u_{\mathrm{rms}}l_f}{\eta}\Bigg[\frac{2\tilde{k}^2}{3(\tilde{k}^4+q^2\tilde{\omega}^2)}\\&+\frac{8(N\tau_c)^2\tilde{k}^2(\tilde{k}^2\frac{Pm^2}{Pr}-q^2\tilde{\omega}^2)}{15(\tilde{k}^4+q^2\tilde{\omega}^2)(\tilde{k}^4Pm^2+q^2\tilde{\omega}^2)(\tilde{k}^4\frac{Pm^2}{Pr^2}+q^2\tilde{\omega}^2)}\Bigg],\nonumber
\end{align}
\begin{align}
    (\tilde{\alpha}^{(0)}_H)_b&=\frac{u_{\mathrm{rms}}l_f}{\eta}\Bigg[-\frac{2\tilde{k}^2Pm}{3(\tilde{k}^4Pm^2+q^2\tilde{\omega}^2)}\nonumber\\&+(N\tau_c)^2\frac{4q^2\tilde{k}^2(\tilde{k}^4\frac{Pm^3}{Pr^2}-(\frac{Pm}{Pr}+2Pm)q^2\tilde{\omega}^2)}{15(\tilde{k}^4Pm^2+q^2\tilde{\omega}^2)^2(\tilde{k}^4\frac{Pm^2}{Pr^2}+q^2\tilde{\omega}^2)}\Bigg],
\end{align}
where $\tilde{k}=kl_f$, $\tilde{\omega}=\omega \tau_c$, $q=l_f^2/\eta\tau_c$ is the ratio of the resistive to the correlation time (a measure conductivity), and $\tau_c$ and $l_f$ are the turbulence correlation time and length.  We note that $(N\tau_c)^2$ is assumed small in the perturbative approach of the SOCA. The formal bound requires that the stratification term be smaller than the dominant terms in the momentum equation, which can be clearly seen by considering the zeroth order momentum and buoyancy equations and balancing the $\partial_tu_{0,x}^{(0)}\sim \theta^{(0)}$ and $\partial_t\theta^{(0)}\sim -N^2u_{0,x}$ terms to get $|u_x^{(1)}|\sim (N\tau_c)^2|u_{0,x}|$. Therefore, $(N\tau_c)^2$ needs to be on the order of the perturbation expansion parameter.

The physical transport coefficients are obtained by an inverse Fourier transform given by  $\beta^{(0)}=(\beta^{(0)})_u+(\beta^{(0)})_b=4\pi\int d\tilde{k} d\tilde{\omega} \tilde{k}^2[ (\tilde{\beta}^{(0)})_u W_u(\tilde{k},\tilde{\omega})+(\tilde{\beta}^{(0)})_bW_b(\tilde{k},\tilde{\omega})]$, where $W_u$ and $W_b$ are the statistics of the non-helical background velocity and magnetic fluctuations, respectively. The non-helical velocity and magnetic fields are assumed to follow Gaussian statistics: $W_u(\tilde{k},\tilde{\omega})=W_b(\tilde{k},\tilde{\omega})=\frac{2\tilde{k}^2}{3(2\pi)^{5/2}}\exp(-\tilde{k}^2/2)/(1+\tilde{\omega}^2)$ \citep{radler2006mean}. The same applies for $\alpha_H^{(0)}$ but using the model statistics for the helical fraction of the turbulence. The isotropic turbulent resistivity here is the same as in the main article ($\eta_t=\beta^{(0)}$) but with a different notation.
\begin{figure}
    \centering
    \includegraphics[width=0.5\linewidth]{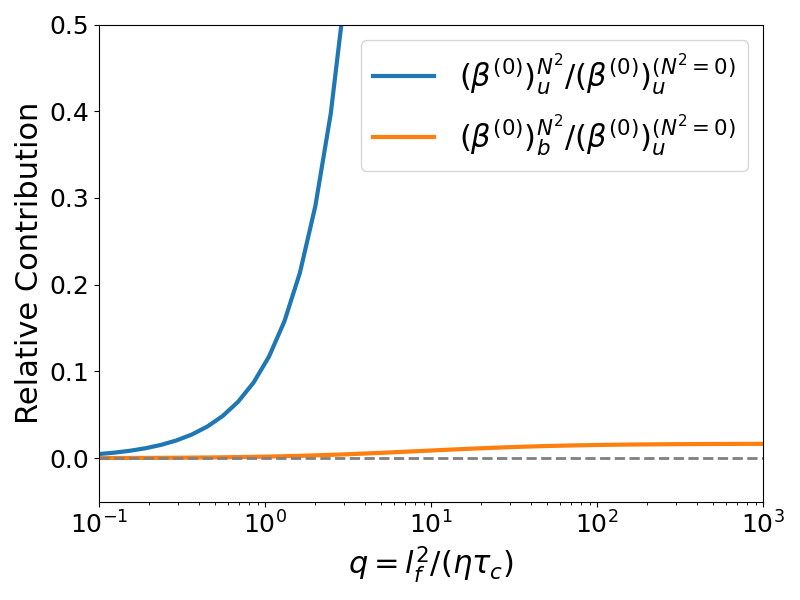}
    \includegraphics[width=0.5\linewidth]{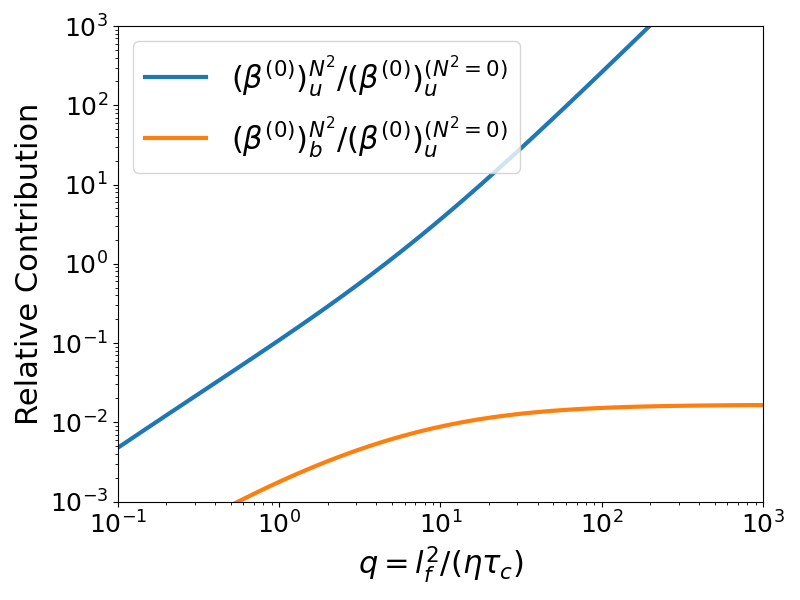}
    \caption{The modification of the isotropic turbulent resistivity by stratification. Plots of the ratio of the stratification term to the unstratified term on a linear (top) and a log scale (bottom). The contribution to the stratification term from velocity fluctuations is in blue and from magnetic fluctuations is in orange. }
    \label{fig:beta}
\end{figure}
We are interested in quantifying the relative size of the stratification term to the unstratified value of $\beta^{(0)}$ since this term affects the MSC effect. To this end, we split the unstratified and stratified contributions to $\beta^{(0)}$ and then examine their ratio. For concreteness, we set unity Prandtl numbers ($Pm=Pr=1$) and define the separate contributions as follows:

\begin{align}
    (\beta^{(0)})_u&=(\beta^{(0)})_u^{(N^2=0)}(q)+(N\tau_c)^2(\beta^{(0)})_u^{N^2}(q),\\ (\beta^{(0)})_b&=(N\tau_c)^2(\beta^{(0)})_b^{N^2}(q),
\end{align}
\begin{align}
    (\beta^{(0)})_u^{(N^2=0)}(q)=4\pi\int d\tilde{k} d\tilde{\omega} \tilde{k}^2\frac{\tilde{k}^2}{3(\tilde{k}^4+q^2\tilde{\omega}^2)}\tilde{W}_u(\tilde{k},\tilde{\omega}),\\
    (\beta^{(0)})_u^{N^2}(q)=4\pi\int d\tilde{k} d\tilde{\omega} \tilde{k}^2\frac{3q^2\tilde{k}^2(\tilde{k}^4-q^2\tilde{\omega}^2)}{10(\tilde{k}^4+q^2\tilde{\omega}^2)^3}\tilde{W}_u(\tilde{k},\tilde{\omega}),\\
    (\beta^{(0)})_b^{N^2}(q)=4\pi\int d\tilde{k} d\tilde{\omega} \tilde{k}^2\frac{q^2\tilde{k}^2(\tilde{k}^4-3q^2\tilde{\omega}^2)}{60(\tilde{k}^4+q^2\tilde{\omega}^2)^3}\tilde{W}_b(\tilde{k},\tilde{\omega}).
\end{align}

Figure \ref{fig:beta} shows the relative size of $(\beta^{(0)})_u^{N^2}(q)$ and $(\beta^{(0)})_b^{N^2}(q)$ with respect to the unstratified value $(\beta^{(0)})_u^{(N^2=0)}(q)$ by plotting their ratios versus $q$, which can be thought of as a measure of the conductivity $q=Rm/St$.  Both kinetic and magnetic contributions from stratification are positive, which is expected since stratification should intuitively reduce the dynamo efficiency. We see that the kinetic contribution is much larger than the magnetic contribution $(\beta^{(0)})_b^{N^2}(q)\gg(\beta^{(0)})_u^{N^2}(q)$. As $q\rightarrow \infty$, the magnetic contribution remains small $\beta^{(0)})_b^{(N^2)}/(\beta^{(0)})_u^{(N^2=0)}\ll1$. The kinetic contribution may become important as $q\rightarrow \infty$ since $(\beta^{(0)})_b^{N^2}(q)/(\beta^{(0)})_u^{(N^2=0)}\gg1$ may become important. However, the formally valid limit $(N\tau_c)^2\ll1$ means that the contributions from stratification are still small compared to the unstratified isotropic turbulent resistivity. While $q\gg1$ seems like the relavent limit for large $Rm$, recall that SOCA is only formally valid for $Rm\ll1$ at $q\ll1$ or $St\ll1$ at $q\gg1$. In reality, $q=O(1)$ probably provides the most reasonable estimate for nonlinear turbulence (see discussions in \citep{brandenburg2005astrophysical,radler2006mean,squire2015electromotive}).

\section{Helicity Generation and Catastrophic Quenching}\label{sec:Helicity}

In this section, we first apply the two-scale approach to the total magnetic helicity, then describe how catastrophic quenching results from helicity constraints, and finally show that the MSC effect does not produce helicity. The resulting takeaway is that the MSC effect is immune to catastrophic quenching and therefore likely remains a robust mechanism in the astrophysical limit of large $Rm$.  

The total magnetic helicity is given by $\mathcal{H}_T=\int \textbf {A}_T\cdot \textbf{B}_T dV$, where $\textbf{A}_T$ is the vector potential ($\textbf{B}_T=\nabla \times \textbf{A}_T$). $\mathcal{H}$ can be split into the helicity of the large scale and small-scale magnetic fields $\mathcal{H}=\int \textbf{A}\cdot \textbf{B}dV$ and $\textit{h}=\int \langle \textbf{a}\cdot \textbf{b}\rangle dV$, where brackets denote the mean field average. Using $\partial_t\textbf{A}_T=-\textbf{E}_T+\nabla\phi$ from the induction equation (where $\textbf{E}_T=\textbf{U}_T\times \textbf{B}_T+\eta \nabla\times \textbf{B}_T$ is the electric field and $\phi$ is an arbitrary scalar field), it is straightforward to show that: 
\begin{align}
    \partial_t\mathcal{H}&=+2\int \mathcal{E}\cdot \textbf{B} dV-2\eta\int (\nabla \times \textbf{B})\cdot \textbf{B}dV,\label{eq:LSHelicity}\\
    \partial_t\textit{h}&=-2\int \mathcal{E}\cdot \textbf{B} dV-2\eta\int \langle (\nabla \times \textbf{b})\cdot \textbf{b}\rangle dV,\label{eq:SSHelicity}
\end{align}
where we have assumed helicity fluxes through the volume boundaries are zero due to periodicity or perfectly conducting boundary conditions. The argument for quenching of helical dynamos stems from the observation that large scale and small scale helicities are produced through the EMF at the same rate, but with opposite signs \citep{rincon2019} (i.e. the source term is $\pm 2\int \mathcal{E}\cdot \textbf{B} dV$). Consider a turbulent MHD system at large $Rm$ with a growing, but still weak mean magnetic field. The SSD will be the first to saturate and so the small-scale helicity will remain roughly constant $\partial_t \textit{h}\approx0$, leaving a balance between between helicity injection and dissipation at small scales in Equation \eqref{eq:SSHelicity} (i.e. $2\int \mathcal{E}\cdot \textbf{B} dV\approx -2\eta\int \langle (\nabla \times \textbf{b})\cdot \textbf{b}\rangle dV$). As a result, the large-scale helicity in Equation \eqref{eq:LSHelicity} becomes constrained to grow at the dissipation rate of small-scale helicity, $\partial_tH\approx-2\eta\int \langle (\nabla \times \textbf{b})\cdot \textbf{b}\rangle dV$ (since dissipation of helicity by the small-scale fields is much faster than by the large-scale fields). The dependence of $\partial_t\mathcal{H}$ on the microscopic resistivity $\eta$ leads to a resistivity limited growth (catastrophic quenching) of the large-scale dynamo in the astrophysical limit of $\eta\rightarrow 0$ (i.e. $Rm\rightarrow \infty$). It may be possible to avoid catastrophic quenching if small-scale helicity can be transported out of the domain boundaries at the production rate of large-scale helicity by dropping the ideal boundary condition assumption. However, simulations have so far found mixed results \citep{rincon2019}.

LSDs driven by helical turbulence produce helical large-scale fields because the form of the EMF gives a non-zero source term. For example, the simplest alpha dynamo $\mathcal{E}_i=\alpha^{(0)}_{H} \textbf{B}_i$ has $\int \mathcal{E}\cdot \textbf{B} dV=\alpha^{(0)}_{H}\int |\textbf{B}|^2 dV\neq0$ and $\partial_t \mathcal{H}\neq0$. The same can be shown for all other alpha-effect based dynamos (e.g. $\alpha$-$\Omega$ dynamos) whose transport coefficients (such as $\alpha^{(0)}_{H}$, $\alpha^{(D)}_{H}$ etc.) are based on the helical part of the background turbulence. 

On the other hand, LSDs driven by non-helical turbulence generate non-helical large scale fields because the source term is zero. The MSC effect is an example of the general class of shear-current effects that have contributions from the following terms in the EMF: 
\begin{align}
\mathcal{E}^{\mathrm{SC}}_i=&-\beta^{(D)}D_{ij}\textbf{J}_j-\delta^{(W)}\epsilon_{ijk}W_j\textbf{J}_{k}-\kappa^{(W)}W_j(\nabla \textbf{B})_{ji}^{(s)}\nonumber\\&-2\kappa^{(D)}\epsilon_{ijk}D_{kr}(\nabla \textbf{B})_{jr}^{(s)}.
\end{align}

The crucial coefficient in the MSC effect, $\eta_{yx}=-\delta^{(W)}+\frac{1}{2}(\kappa^{(W)}-\beta^{(D)}+\kappa^{(D)})$, has contributions from each of these terms. It is straightforward to show that $\int \mathcal{E}^{\mathrm{SC}}\cdot \textbf{B} dV=0$, with each term independently having a null contribution. Consider the term proportional to $\kappa^{(W)}$:

\begin{align}
\int \kappa^{(W)}W_j(\nabla \textbf{B})_{ji}^{(s)}B_idV=\frac{1}{2}\kappa^{(W)}W_j\int \left(\frac{\partial \textbf{B}_j}{\partial x_i}+\frac{\partial \textbf{B}_i}{\partial x_j}\right)B_idV=\\=\frac{1}{2}\kappa^{(W)}W_j\int \left(\frac{\partial (B_jB_i)}{\partial x_i}+\frac{\partial (B_iB_i)}{\partial x_j}\right)dV=0 ,   
\end{align}
where the divergence free condition has been used, $\frac{\partial B_i}{\partial x_i}=0$, and we have assumed either periodic boundary conditions or that fields vanish sufficiently fast outside a finite region. Closed boundary conditions could in principle have a non-zero contribution \citep{brandenburg2005BC_helicityflux}. The calculation for every other term is similar. As a result, the MSC effect, driven by non-helical magnetic fluctuations, generates non-helical large-scale magnetic fields and is not affected by the helicity constraints usually used to argue for the inevitability of catastrophic quenching at high $Rm$.

\section{Long Term Behavior of the Saturated Large-Scale Dynamo}\label{sec:LongRun}
\begin{figure}
    \centering
    \includegraphics[width=0.7\linewidth]{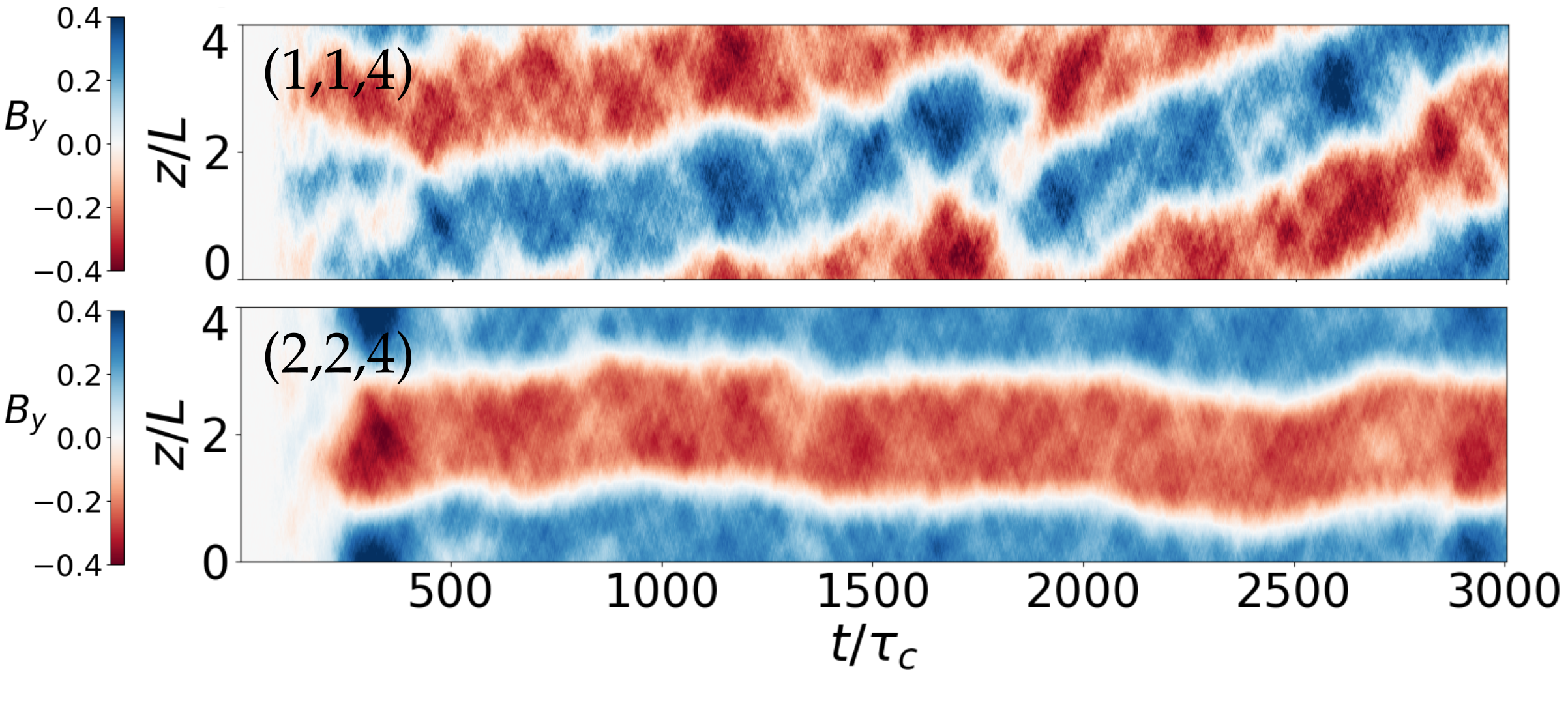}
    \caption{A test showing that the long-term trend of the saturated LSD phase becomes more coherent with increasing domain size where incoherent effects become less efficient. Time-space plots from single realizations of the y-component of the magnetic field $B_y(z,t)=\langle B_y(x,y,z,t)\rangle_{x,y}$ for two aspect ratios denoted in the top left corner of each plot are shown. Both simulations have fixed values of $Rm\approx60$, $Sh\approx1.0$, and $Fr^{-1}\approx0.2$ and the spectral resolution is scaled with the aspect ratio. The fiducial $(1,1,4)$ simulation has $N_x\times N_y\times N_z=96^2\times384$ modes.   }
    \label{fig:LongRuns}
\end{figure}

This section briefly shows the dependence of the long term evolution of the saturated LSD on the domain size. Viewing the dynamo as an instability, fluctuations of the large scale magnetic field at the end of the rapid SSD phase will grow at rates given by the dispersion relation of any present LSD instabilities, which in this setup is a possible combination of the coherent MSC effect and incoherent effects. As discussed in Section \ref{sec:IncoherentTest}, the incoherent effects become less efficient with increasing volume while coherent effects are volume independent (if the dominant mode fits in the domain). Figure \ref{fig:LongRuns} compares the evolution of the saturated LSD across a long time scale $t/\tau_c=3000$ (compared to the measured growth period of the LSD, $\gamma_{\rm LSD}^{-1}\sim 20\tau_c$ from Table \ref{tab:LSDgrowthRates}) between the two cases with domain sizes (1,1,4) and (2,2,4). The (4,4,4) case is not included because it was too expensive to run for long times.

The long term evolution of the (1,1,4) case shown in the top panel of Figure \ref{fig:LongRuns} has a constant phase for the first $t/\tau_c \sim 1000$ but then begins to very slowly vary on $t/\tau_c \sim 500$ timescales. This may be related to the non-linear saturation mechanism of the MSC effect, an incoherent effect, or an interaction of the two. When the volume is quadrupled in the (2,2,4) case, any incoherent effect is expected to become suppressed by a factor of 2. As shown in the bottom panel of Figure \ref{fig:LongRuns}, the (2,2,4) case maintains a relatively steady phase across the entire duration as might be expected from the saturation of the coherent MSC effect alone (whose growth rate is purely real). This suggests that the domain volume is now sufficiently large that incoherent effects have become insignificant.

\bibliography{references}{}
\bibliographystyle{aasjournal}



\end{document}